\documentclass[sigconf,screen]{acmart}

\AtBeginDocument{%
  }

\copyrightyear{2026}
\acmYear{2026}
\setcopyright{cc}
\setcctype{by}
\acmConference[ASE '26]{Proceedings of the 41st IEEE/ACM International Conference on Automated Software Engineering}{October 12--16, 2026}{Munich, Germany}
\acmBooktitle{Proceedings of the 41st IEEE/ACM International Conference on Automated Software Engineering (ASE '26), October 12--16, 2026, Munich, Germany}
\acmDOI{10.1145/3832783.3834361}
\acmISBN{979-8-4007-2882-2/2026/10}
\ifdefined\ATLASCameraReady
  \makeatletter
  \def\@authorfont{\large}
  \def\@affiliationfont{\small}
  \makeatother
\else
\fi

\usepackage{xcolor}
\usepackage{tikz}
\usetikzlibrary{shapes,shapes.geometric,arrows,positioning,fit,calc,backgrounds}
\usepackage{forest}
\usepackage{booktabs}
\usepackage{algorithm}
\usepackage{algorithmic}
\usepackage{colortbl}
\usepackage{array}
\usepackage{subcaption}
\usepackage{balance}
\usepackage{fontawesome5}
\usepackage[most]{tcolorbox}

\definecolor{cbBlue}{HTML}{0072B2}
\definecolor{cbOrange}{HTML}{E69F00}
\definecolor{cbGreen}{HTML}{009E73}
\definecolor{cbRed}{HTML}{D55E00}
\definecolor{cbPurple}{HTML}{CC79A7}
\definecolor{loopBg}{HTML}{F5F9FC}
\definecolor{cbYellow}{HTML}{F0E442}

\newcommand{\aitd}{ATLAS}
\newcommand{\designeragent}{Designer Agent}
\newcommand{\classifieragent}{Classifier Agent}
\newcommand{\finding}[2]{%
  \smallskip
  \begin{tcolorbox}[
    colback=gray!8,
    colframe=gray!40,
    boxrule=0.4pt,
    left=4pt, right=4pt, top=3pt, bottom=3pt,
    arc=1.5pt, outer arc=1.5pt,
    width=\columnwidth,
  ]
  \textbf{Finding #1:} #2
  \end{tcolorbox}
  \smallskip
}
\newcommand{\observation}[2]{\noindent\textbf{Observation #1: #2}
}
\begin{document}

\title[ATLAS]{ATLAS: Agentic Taxonomy of LArge-Scale Software Ecosystems}

\author{Junyi Lu}
\affiliation{%
  \institution{Institute of Software, Chinese Academy of Sciences; University of Chinese Academy of Sciences}
  \city{Beijing}
  \country{China}}
\email{lujunyi21@mails.ucas.ac.cn}

\author{Mengyao Lyu}
\authornote{Corresponding authors.}
\affiliation{%
  \institution{Nanyang Technological University}
  \city{Singapore}
  \country{Singapore}}
\email{mengyao.lyu@ntu.edu.sg}

\author{Jiahui Wu}
\affiliation{%
  \institution{Nanyang Technological University}
  \city{Singapore}
  \country{Singapore}}
\email{jiahui004@e.ntu.edu.sg}

\author{Lei Yu}
\affiliation{%
  \institution{Institute of Software, Chinese Academy of Sciences; University of Chinese Academy of Sciences}
  \city{Beijing}
  \country{China}}
\email{yulei2022@iscas.ac.cn}

\author{Chengwei Liu}
\affiliation{%
  \institution{Nankai University}
  \city{Tianjin}
  \country{China}}
\email{chengwei.liu@nankai.edu.cn}

\author{Fengjun Zhang}
\affiliation{%
  \institution{Institute of Software, Chinese Academy of Sciences}
  \city{Beijing}
  \country{China}}
\email{fengjun@iscas.ac.cn}

\author{Li Yang}
\authornotemark[1]
\affiliation{%
  \institution{Institute of Software, Chinese Academy of Sciences}
  \city{Beijing}
  \country{China}}
\email{yangli2017@iscas.ac.cn}

\author{Chun Zuo}
\affiliation{%
  \institution{Sinosoft Company Limited}
  \city{Beijing}
  \country{China}}
\email{zuochun@sinosoft.com.cn}

\author{Yang Liu}
\affiliation{%
  \institution{Nanyang Technological University}
  \city{Singapore}
  \country{Singapore}}
\email{yangliu@ntu.edu.sg}
\renewcommand{\shortauthors}{Lu et al.}

\begin{abstract}
The open-source ecosystem on GitHub lacks a systematic hierarchical
taxonomy of software repositories.
GitHub Topics, the dominant organizational mechanism, is flat,
inconsistent, and covers only 67\% of repositories.
We present ATLAS, the first framework that automatically constructs
a hierarchical taxonomy for software repositories and classifies
repositories into it end-to-end.
By combining LLM global knowledge with real repository distributions,
ATLAS proposes meaningful splitting dimensions and iteratively corrects
those that fail to accommodate real repositories.
A Designer Agent proposes splitting dimensions while a Classifier
Agent assigns repositories; a self-corrective refinement loop uses
classification failures to drive dimension revision through
escalating strategies.
We evaluate ATLAS on 54,387 GitHub repositories against six baselines
spanning four paradigm families, two downstream software engineering
tasks, and three model families.
On a stratified 2{,}001-repository benchmark, ATLAS achieves a
Taxonomy Quality F-score (TQF) of 83.13\%,
outperforming the best baseline by 15 percentage points
(on the full 54k corpus the approximate TQF is 73.0\%, a gap driven
by Path Granularity's all-or-nothing scoring on longer paths rather
than lower classification accuracy; see Section~\ref{sec:rq3}).
It is the only method to simultaneously achieve high structural
quality and high practical applicability.
On downstream tasks, the ATLAS taxonomy enables software alternative
discovery with P@1 = 85.71\%, surpassing even human-curated lists
(62.34\%), and achieves the highest P@1 for repository retrieval.
The taxonomy further reveals structural ecosystem trends
that are difficult to obtain from flat tags or similarity methods:
the shift from libraries to AI/ML applications (now 61\% of
newly community-adopted projects) becomes visible only through
hierarchical, type-based categorization.
An interactive taxonomy explorer is available at
\url{https://atlas-taxonomy.netlify.app/}.
\end{abstract}



\maketitle

\section{Introduction}
\label{sec:intro}

When a developer searches for an alternative to Kubernetes for
container orchestration, GitHub returns thousands of repositories
tagged \texttt{kubernetes} (tools, documentation, extensions, and
tutorials) but virtually no actual orchestration alternatives.
The root cause is not the search algorithm but the lack of a
systematic \emph{taxonomy}.
GitHub Topics, the ecosystem's only organizational mechanism, assigns
flat, user-defined tags with no hierarchy, no normalization
(e.g., \texttt{machine-learning} vs.\ \texttt{ml}), and no
coverage guarantee; in our dataset, 33\% of repositories with
$\geq$1,000 stars lack topics.
The result is over 54,000 community-adopted repositories with no systematic way to navigate, compare, or analyze.

What the ecosystem needs is a \emph{taxonomy}: a hierarchical
system where every category has a precise definition, sibling
categories are mutually exclusive, and tree depth encodes increasing
specificity~\cite{mayr1969principles}.
A well-constructed taxonomy enables capabilities that flat tags
cannot: \emph{multi-granularity retrieval} (searching at any level
of specificity), \emph{alternative discovery} (finding functionally
equivalent projects as siblings in the same category), and
\emph{ecosystem analysis} (revealing structural trends across
domains and time).
Crucially, these capabilities require \emph{structured} hierarchies
with explicit definitions and splitting criteria, not merely tag
matching or embedding similarity, which lack hierarchical
structure and cannot distinguish a project's functional role from its
topical association.
The primary beneficiaries are developers who search for, compare, and
retrieve repositories by what they do, with ecosystem analysts a
secondary audience; we make this intended use and its scope precise in
Section~\ref{sec:scope}.
Prior work has organized software domain \emph{terms} into
IS-A ontologies~\cite{sas2024bottomup,sas2023gitranking} that link
hundreds of curated labels via hypernym relations drawn from
external knowledge sources (a fundamentally different structure from
a classification taxonomy whose categories are defined by splitting
dimensions over actual items), or classified repositories into
\emph{predefined} category
schemes~\cite{zhang2019higitclass}.
No prior work has automatically constructed a data-driven
classification taxonomy at scale and classified repositories into
it end-to-end.

Existing approaches each address only part of this problem, leaving
critical gaps.
LLM-based taxonomy methods such as
Chain-of-Layer~\cite{zeng2024col} build taxonomies from pure LLM
knowledge without grounding in actual data, producing well-organized
structures that nevertheless fail to accommodate real repositories.
This is a \emph{depth} problem: the right \emph{splitting dimension}
(the conceptual axis along which categories are defined) at each
level depends on what items actually exist, not what a language model
expects.
Single-pass designs inevitably miss edge cases: repositories that
defy clean categorization expose coverage gaps and definition
ambiguities that no sample-based design can fully anticipate,
a \emph{breadth} problem.
Embedding-based clustering groups repositories by pairwise similarity
but lacks the top-down semantic knowledge to choose meaningful
splitting criteria at each level: it can detect \emph{that} items
are similar, but not articulate \emph{why} they should be grouped
or how siblings differ, an \emph{interpretability} problem.
Beyond these per-method limitations, none of these approaches
\emph{scale}: LLM-based methods face context-window and cost limits
that prevent processing tens of thousands of items, single-pass
designs accumulate uncorrected errors as the collection grows, and
embedding clustering produces increasingly unmanageable flat
partitions.
Effective taxonomy construction therefore requires two ingredients:
the semantic knowledge that LLMs provide (to choose principled
splitting dimensions) and grounding in actual data distributions
(to ensure categories accommodate real repositories).
It further requires an architecture that processes the taxonomy
incrementally, node by node, so that cost and complexity grow with
tree size rather than exploding with corpus size.
The key insight behind our approach is that taxonomy design must be
\emph{self-corrective}: classification failures are not noise to be
discarded, but diagnostic signals that reveal how the current design
is deficient and must feed back into dimension revision.

\begin{figure*}[t]
  \centering
  \ifdefined\ATLASCameraReady
    \includegraphics[width=\textwidth]{figures/fig1_camera_ready.pdf}
  \else
    \includegraphics[width=\textwidth]{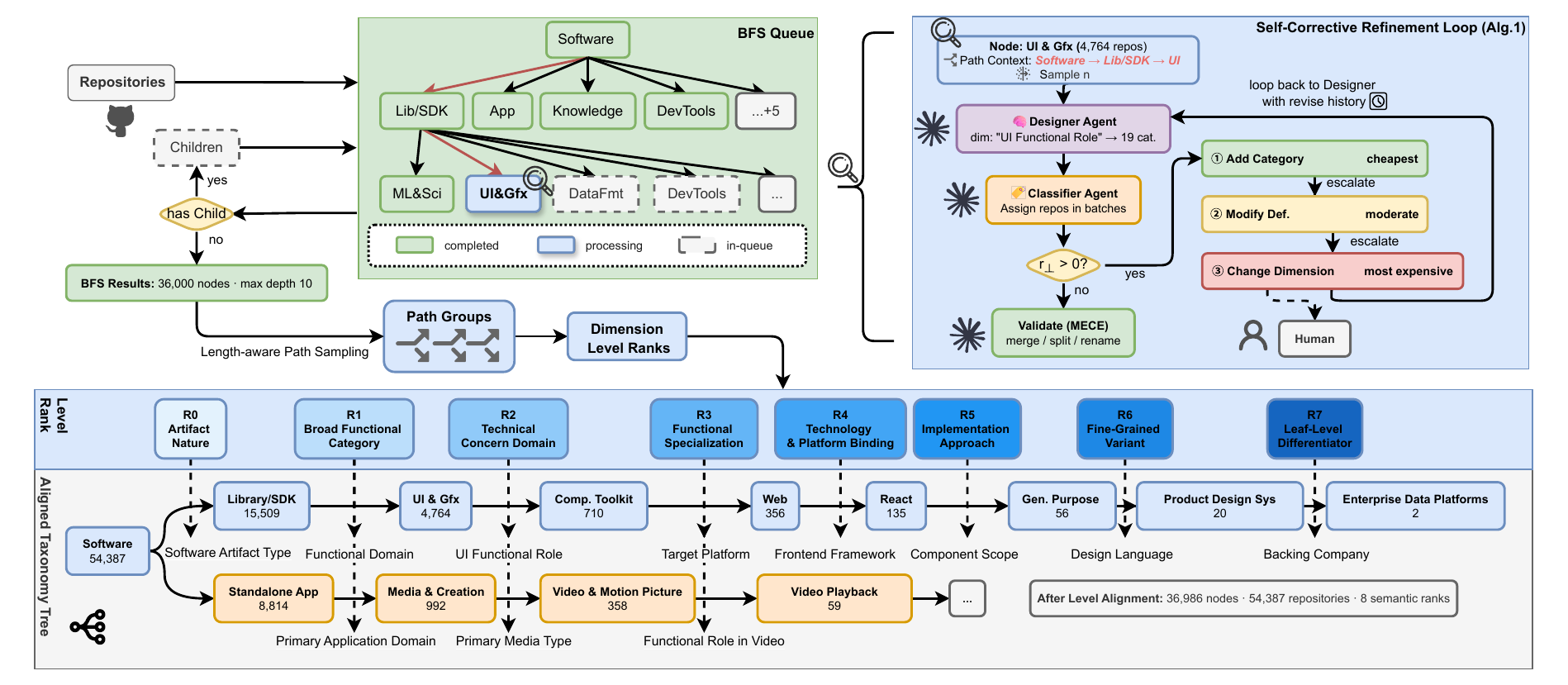}
  \fi
  \caption{Overview of the \aitd{} framework.
  \emph{Top left:} \aitd{} constructs the taxonomy through breadth-first top-down traversal, processing one node at a time.
  \emph{Top right:} At each node, the \designeragent{} proposes a splitting dimension, the \classifieragent{} assigns repositories, and classification failures trigger self-corrective refinement through three escalating strategies; a MECE (mutually exclusive, collectively exhaustive) validation step checks the result before proceeding.
  \emph{Bottom:} The resulting taxonomy of 54,387 repositories spans 36,986 nodes across 8 semantic ranks (R0--R7). Two paths are highlighted: a deep path (blue) from root to a leaf design-system category through 8 ranks, and a shallow path (orange) ending at rank~4, showing that different domains need different levels of specificity.}
  \Description{Architecture diagram showing the ATLAS framework: top-left shows BFS tree traversal, top-right shows the single-node self-corrective refinement loop with Designer Agent, Classifier Agent, and validation, bottom shows the constructed taxonomy with deep and shallow path examples across 8 semantic ranks (R0--R7).}
  \label{fig:architecture}
\end{figure*}

We present \aitd{} (Figure~\ref{fig:architecture}), a multi-agent
framework that addresses these challenges through a
design-classify-refine architecture.
\aitd{} separates taxonomy \emph{design} (a \designeragent{} that
proposes splitting dimensions grounded in repository samples) from
\emph{classification} (a \classifieragent{} that assigns
repositories), connected by a self-corrective refinement loop where
classification failures drive iterative dimension revision.
The \designeragent{} draws on LLM global knowledge to identify
semantically meaningful splitting dimensions, then grounds them in
repository samples to ensure dimensions fit actual data
distributions (\emph{depth}); the refinement loop catches coverage
gaps and definition ambiguities that any single design pass would
miss (\emph{breadth}); and explicit dimension names,
category definitions, and splitting rationale at every node produce
an interpretable taxonomy without post-hoc labeling
(\emph{interpretability}).
This separation also enables asymmetric model pairing (a stronger
model for design, a weaker model for classification) and
breadth-first node-by-node traversal where each node only requires
a compressed path context (the ancestor chain of dimensions) rather
than the full tree, making cost and complexity scale with tree size
rather than corpus size, addressing \emph{scale}.
After construction, a Level Alignment phase establishes a unified
semantic rank system across branches, ensuring cross-branch
granularity consistency.\looseness=-1

We evaluate \aitd{} on 54,387 GitHub repositories ($\geq$1,000
stars), producing the largest automated software taxonomy to date.
On a stratified 2{,}001-repository benchmark, \aitd{} achieves a
Taxonomy Quality F-score (TQF) of 83.13\%,
outperforming the best baseline by 15 percentage points
(the approximate TQF on the full 54k corpus is 73.0\%, a gap driven
by Path Granularity's all-or-nothing scoring on longer paths rather
than lower classification accuracy; Section~\ref{sec:rq3}).
It is the only method that simultaneously attains high structural
quality (TQ $>$ 75\%) and high practical applicability
(NCP $>$ 90\%).
On downstream software engineering tasks, \aitd{}'s taxonomy
enables alternative discovery with P@1 = 85.71\%, outperforming
tag-based (55.56\%) and embedding (61.69\%) approaches and
surpassing even human-curated lists (62.34\%).
This precision arises because taxonomy leaf categories, defined by
explicit splitting dimensions, group repositories by functional
equivalence rather than loose topical association.
The taxonomy also provides a broader lens on the open-source
ecosystem, revealing structural trends that are difficult to obtain
from flat tags or similarity methods, such as the shift from
libraries to AI/ML applications (now 61\% of newly community-adopted
projects).

Our contributions are:
\begin{itemize}
  \item We formulate the problem of end-to-end taxonomy construction
    and classification for software repositories and, to the best of
    our knowledge, present the first framework addressing this task
    at scale.
  \item We propose \aitd{}, a multi-agent architecture that combines
    data-driven dimension design, batch classification with diagnostic
    feedback, iterative self-corrective refinement, MECE (mutually
    exclusive, collectively exhaustive) validation,
    and dimension-based level alignment.
  \item We introduce TQF, a composite metric that balances structural
    quality with practical applicability, addressing the need for
    holistic taxonomy evaluation.
  \item We evaluate \aitd{} on 54,387 repositories against six
    baselines and two downstream tasks, with ablation and cross-model
    studies spanning three model families.
\end{itemize}

\section{Background and Motivation}
\label{sec:background}

\subsection{Taxonomy vs.\ Folksonomy}
\label{sec:bg-taxonomy}

Section~\ref{sec:intro} defined a taxonomy as a hierarchical
classification system with precise definitions and mutual exclusivity
at each level, contrasting it with GitHub Topics' flat
folksonomy~\cite{vander2007folksonomies,izadi2021topic}.
The value of such hierarchies is well established in
biology~\cite{mayr1969principles}, but software repositories pose
unique challenges compared to text corpora.
Items span diverse artifact types (libraries, applications, tools,
frameworks), technical stacks, and functional domains.
This heterogeneity demands domain-aware \emph{dimension design} at
each level rather than generic topic clustering.
Existing taxonomy methods~\cite{zeng2024col,taxoadapt2025} have not
been applied to this setting (Section~\ref{sec:related}).

\subsection{Motivating Scenario}
\label{sec:bg-motivation}

Returning to the Kubernetes scenario from Section~\ref{sec:intro}:
GitHub Topics returns K8s \emph{tools} (\texttt{kops}, \texttt{minikube}), K8s \emph{documentation} (\texttt{kubernetes\allowbreak{}-in\allowbreak{}-action}), and K8s \emph{extensions} (\texttt{kruise}), none of which are orchestration alternatives.
Embedding similarity retrieves projects with similar descriptions
but different purposes (e.g., \texttt{kubesphere}, a K8s management
platform).
Neither approach can distinguish a project's \emph{functional role}
from its \emph{topical association}.

An \aitd{}-constructed taxonomy resolves this by classifying
Kubernetes under \textit{Platform/Service $\to$ Container Orchestration},
with true alternatives (\texttt{mesos}, \texttt{dcos},
\texttt{service-fabric}) as siblings in the same leaf category.
This precision arises from two properties: the taxonomy defines
categories with explicit \emph{splitting dimensions}
(e.g., ``by deployment model'') that separate functional roles,
and the iterative construction process refines categories based on
actual classification outcomes.
Fig.~\ref{fig:architecture} illustrates how \aitd{} produces such
a taxonomy through its multi-agent iterative framework.

\subsection{Intended Stakeholders and Scope}
\label{sec:scope}

\aitd{} provides a \emph{general functional organization} of software
repositories that is comparable in scope to GitHub Topics, yet
hierarchical and definition-grounded.
Its primary stakeholders are developers who search for, compare, and
retrieve repositories by what they do, and three properties of the
taxonomy address their needs.
First, for \emph{finding true substitutes}, leaf categories are defined
by explicit splitting dimensions that group repositories by functional
role rather than topical association, so a developer replacing one tool
retrieves genuine alternatives; this yields P@1 = 85.71\% on
alternative discovery, above a human-curated list
(Section~\ref{sec:alt-discovery}).
Second, for \emph{searching at the right granularity}, the multi-rank
taxonomy with per-node definitions lets a developer enter at a coarse
category and drill down, giving the highest P@1 on retrieval
(Section~\ref{sec:retrieval}).
Third, for \emph{interpretable results}, every category carries a name,
definition, and splitting rationale, so a developer can see \emph{why}
two projects are or are not alternatives instead of relying on an
opaque similarity score.
A secondary stakeholder is the ecosystem analyst, for whom the
hierarchical, type-based structure (the L1 artifact-type axis) reveals
structural trends---such as the library-to-AI/ML shift---that flat tags
or embedding clusters do not expose (Section~\ref{sec:insights}).

\aitd{} organizes repositories by function, and other goals call for
different facets.
Assessing maintenance risk, license compatibility, or security posture
draws on activity, license, or vulnerability axes that a functional
taxonomy does not encode, and these fall outside our scope.
The Dimension Stability Principle (Section~\ref{sec:designer})
prioritizes stable, essential properties---artifact type and functional
domain---that capture a repository's enduring functional role, and
because the \designeragent{}'s prompt determines which dimension each
node uses, the same architecture can target a different facet when one
is needed.

\section{The ATLAS Framework}
\label{sec:approach}

This section details each component of \aitd{}: dimension design (\S\ref{sec:designer}), batch classification (\S\ref{sec:classifier}), iterative refinement (\S\ref{sec:refinement}), single-layer validation (\S\ref{sec:validation}), and level alignment (\S\ref{sec:alignment}).

\subsection{Overview}
\label{sec:overview}

Constructing a taxonomy is a \emph{schema design} problem: at each level, the system must choose the right conceptual axis, what we call a \emph{splitting dimension}, along which to divide repositories into categories.
\aitd{} treats dimension design as the central challenge and classification as a subordinate execution step.

As introduced in Section~\ref{sec:intro}, the framework separates taxonomy construction into a \textbf{\designeragent{}} (dimension design) and a \textbf{\classifieragent{}} (repository assignment), connected by a self-corrective refinement loop (Figure~\ref{fig:architecture}).
The construction follows breadth-first, top-down traversal: at each node, \aitd{} runs a \emph{single-layer loop} (Algorithm~\ref{alg:single-layer}) of design, classification, and refinement, followed by validation before child nodes are enqueued.
Top-down construction ensures every node receives explicit names, definitions, and splitting rationale, producing an interpretable taxonomy, while BFS enables level-based checkpointing and parallelization at scale.

\subsection{Dimension Design}
\label{sec:designer}

The central challenge in taxonomy construction is choosing the right splitting dimension at each node.
The \designeragent{} must make this decision based on a \emph{local} view (a sample of the node's repositories) while maintaining \emph{global} coherence with the rest of the tree.
Two mechanisms address this tension.

\paragraph{Path Context.}
The complete taxonomy tree cannot fit in a single LLM context window, so we provide the \designeragent{} with a \emph{path context}: the full chain of ancestor node names, definitions, and splitting dimensions from root to current node.
This compressed representation of the tree-so-far serves two purposes.
First, it prevents redundant dimensions: the \designeragent{} can see that ``functional domain'' was already used at depth~1 and will not repeat it at depth~3.
Second, it enforces a coarse-to-fine progression: the \designeragent{} is instructed to choose dimensions that are \emph{more specific} than those of its ancestors, ensuring granularity increases monotonically down the tree.

\paragraph{Dimension Stability Principle.}
Not all dimensions are equally suitable for upper levels of the taxonomy.
Borrowing from biological taxonomy, where higher ranks (kingdom, phylum) are far more stable than lower ones (genus, species)~\cite{mayr1969principles}, we require that upper-level dimensions capture \emph{stable, essential features} (such as functional domain or artifact type) that are less likely to change as the software ecosystem evolves.
Lower-level dimensions may use more volatile features such as target platform or implementation approach.
This principle aims to keep the taxonomy's upper-level structure stable as the ecosystem evolves: adding new repositories may expand leaf categories, but the fundamental axes persist.

Given these mechanisms, the \designeragent{} receives a random sample of the node's repositories (80 repositories, $\approx$22k tokens, balancing context-window capacity with LLM attention limits) along with the path context, and produces a \emph{dimension design}: a dimension description, a set of child categories each with a name, definition, and inclusion criteria, and a rationale for the choice.

\subsection{Batch Classification}
\label{sec:classifier}

The \classifieragent{} assigns repositories to the categories defined by the \designeragent{}, processing them in batches to amortize the cost of including the full dimension design in each prompt.
For each repository, it produces an assigned category (or \texttt{null} if none fits), a confidence score, and a brief reasoning.

The distinction between \texttt{null} and low-confidence assignments enables targeted refinement.
A \texttt{null} assignment indicates a \emph{coverage gap}: the taxonomy is missing an appropriate category for the repository.
A low-confidence assignment indicates \emph{definition ambiguity}: category boundaries overlap, making the correct placement uncertain.
This diagnostic signal lets the refinement loop (\S\ref{sec:refinement}) target specific structural weaknesses rather than operating blindly.

\subsection{Iterative Refinement}
\label{sec:refinement}

A single round of dimension design rarely produces a perfect category set, because the \designeragent{} works from a sample rather than the full repository set.
Classification failures (repositories that receive \texttt{null} assignments) reveal how the current design is deficient.
The key insight is that \emph{the pattern of failures contains diagnostic information}.
Given a batch $B$ classified under dimension design $D$, let $f(x, D)$ denote the category assigned to repository $x$ under $D$ ($\bot$ if no category fits).
We define the \emph{null rate}:
\begin{equation}
  r_\bot(B, D) = \frac{|\{x \in B \mid f(x, D) = \bot\}|}{|B|}
\end{equation}
Ideally $r_\bot = 0$ (all repositories classified); any $r_\bot > 0$ triggers refinement.
The magnitude is diagnostic: a low $r_\bot$ points to localized coverage gaps (a few missing categories), while a high $r_\bot$ suggests the splitting dimension itself is unsuitable for the data.
The system addresses failures through three strategies ordered by increasing disruption (expanding the category set, refining definitions, or replacing the dimension entirely), following a \emph{minimum-intervention principle}: start with the lightest fix and escalate only when it fails to reduce $r_\bot$.
To avoid repeating failed approaches, the \designeragent{} receives a \emph{modification history}, the sequence of previous revision types and the before/after category changes for each attempt, enabling informed strategy selection across iterations.

\smallskip
\noindent\textbf{(1) Add Category.}
Introduce new categories for unclassified repositories.
Only the failed repositories are reclassified, making this the cheapest strategy.
Appropriate when the dimension is sound but the initial category set was incomplete.

\smallskip
\noindent\textbf{(2) Modify Definition.}
Clarify or adjust category definitions to resolve boundary ambiguity.
Failed and displaced repositories are reclassified against the updated design.
Appropriate when repositories fall between categories due to imprecise definitions.

\smallskip
\noindent\textbf{(3) Change Dimension.}
When lighter strategies repeatedly fail to reduce $r_\bot$, the system replaces the entire splitting dimension.
All repositories are reclassified from scratch, making this the most expensive strategy.

\smallskip
We denote these strategies $s_1$ (Add Category), $s_2$ (Modify Definition), $s_3$ (Change Dimension) in order of increasing disruption.
After applying $s_i$, if $r_\bot$ does not decrease ($r_\bot^{(t)} \geq r_\bot^{(t-1)}$), the system escalates to $s_{i+1}$.
A \emph{downgrade rule} prevents premature escalation:
\begin{equation}
  s^{*} = \begin{cases}
    s_1 \text{ (Add Category)} & \text{if } s = s_3 \text{ and } r_\bot < 0.5 \\
    s & \text{otherwise}
  \end{cases}
\end{equation}
That is, localized failures ($r_\bot < 0.5$) are addressed by expanding the category set rather than discarding the entire dimension.
For large nodes, an \emph{early revision trigger} interrupts classification as soon as the running null rate exceeds a dynamic threshold, avoiding thousands of wasted LLM calls under a flawed dimension.
Algorithm~\ref{alg:single-layer} formalizes the complete single-layer loop.

\begin{algorithm}[t]
\caption{Single-Layer Processing Loop}
\label{alg:single-layer}
\begin{algorithmic}[1]
\REQUIRE Node $n$ with repositories $R$, max revisions $k$
\ENSURE Child nodes $\{c_1, c_2, \ldots\}$
\STATE $S \gets \textsc{RandomSample}(R)$
\STATE $D \gets \textsc{DesignerAgent.Design}(S, n.\mathit{pathCtx})$
\STATE $\mathit{children} \gets \textsc{CreateNodes}(D.\mathit{categories})$
\STATE $\mathit{ok}, \mathit{fail} \gets \textsc{ClassifyWithEarlyTrigger}(R, D)$
\STATE $\mathit{hist} \gets []$ \COMMENT{modification history}
\FOR{$i = 1$ \TO $k$ \textbf{while} $\mathit{fail} \neq \emptyset$}
    \STATE $\mathit{strat} \gets \textsc{DesignerAgent.Revise}(\mathit{fail}, D, \mathit{hist})$
    \IF{$|\mathit{fail}| / |R| < 0.5$ \textbf{and} $\mathit{strat} = \textsc{ChangeDim}$}
        \STATE $\mathit{strat} \gets \textsc{AddCat}$ \COMMENT{downgrade}
    \ENDIF
    \STATE $D_{\mathrm{old}} \gets D$
    \IF{$\mathit{strat} = \textsc{ChangeDim}$}
        \STATE $D \gets \textsc{DesignerAgent.Design}(S, n.\mathit{pathCtx})$
        \STATE $\mathit{ok}, \mathit{fail} \gets \textsc{Classify}(R, D)$ \COMMENT{all repositories}
    \ELSE
        \STATE $D \gets \textsc{UpdateDesign}(D, \mathit{strat}, \mathit{fail})$
        \STATE $a, f \gets \textsc{Classify}(\mathit{fail}, D)$ \COMMENT{affected only}
        \STATE $\mathit{ok} \gets \mathit{ok} \cup a$; \quad $\mathit{fail} \gets f$
    \ENDIF
    \STATE $\mathit{hist}.\mathrm{append}(\mathit{strat}, D_{\mathrm{old}}, D)$
\ENDFOR
\STATE \textsc{AssignRepos}($\mathit{children}$, $\mathit{ok}$)
\STATE \textsc{RemoveEmptyCategories}($\mathit{children}$)
\STATE $\mathit{children} \gets \textsc{Validate}(\mathit{children})$ \COMMENT{merge/split/rename}
\IF{$\mathit{fail} \neq \emptyset$}
    \STATE Route $\mathit{fail}$ to human review queue
\ENDIF
\RETURN $\mathit{children}$
\end{algorithmic}
\end{algorithm}

\subsection{Single-Layer Validation}
\label{sec:validation}

While iterative refinement (\S\ref{sec:refinement}) addresses \emph{classification failures} (repositories that no category fits), validation addresses \emph{structural defects} in the resulting categories before they propagate deeper.

The \designeragent{} examines the child categories along with a sample of classified repositories and checks whether they satisfy the \emph{MECE principle} (Mutually Exclusive, Collectively Exhaustive)~\cite{minto1987pyramid}: category definitions must not overlap, and must collectively cover the repository set without systematic gaps.
Beyond MECE, the validator also checks \emph{distinguishability}: category names and definitions must be sufficiently distinct to avoid confusion.
Notably, the validator does \emph{not} enforce balance; categories may naturally differ greatly in size, and forcing balance would distort the taxonomy's fidelity to the data.
When violations are detected, the validator suggests \textsc{Merge}, \textsc{Split}, or \textsc{Rename} operations, which are applied automatically with affected repositories reassigned.

\subsection{Level Alignment}
\label{sec:alignment}

Because the \designeragent{} makes locally optimal choices at each node, different branches independently select dimensions that serve the same semantic role. For instance, ``UI Functional Role'' in one branch and ``Primary Media Type'' in another both carve out a technical concern domain (R2 in Figure~\ref{fig:architecture}).
Level alignment is a post-construction phase that discovers this implicit structure and establishes a global \emph{dimension rank system}, grouping semantically equivalent dimensions into unified ranks and correcting any depth inconsistencies across branches.

\paragraph{Alignment Unit.}
The unit of alignment is the \emph{splitting dimension}, not the node or category.
Semantic granularity is determined by the sequence of splitting dimensions along a root-to-leaf path: earlier dimensions make coarser distinctions, later ones make finer ones.
Two branches at different depths have equivalent granularity if their dimension sequences cover the same scope.
Aligning dimension paths is thus more fundamental than aligning depths.

\paragraph{Process.}
Level alignment proceeds in three steps:

\begin{enumerate}
    \item \textbf{Rank Discovery.}
    An LLM analyzes a stratified sample of root-to-leaf \emph{dimension paths} and identifies a set of semantic \emph{ranks}, ordered granularity levels such as ``artifact nature'' $\rightarrow$ ``domain category'' $\rightarrow$ ``functional specialty'' $\rightarrow$ ``technical concern.''
    Each dimension is assigned to a semantic rank level via majority voting across all paths that share it, producing a globally consistent mapping.
    Multiple consecutive dimensions may share the same rank (sub-layers) when they refine the same level of specificity.

    \item \textbf{Dimension--Rank Gap Detection.}
    The system checks each branch for gaps (dimensions that jump from a coarse rank to a fine rank without any intermediate step) and for inconsistencies where sibling nodes were produced by dimensions at different ranks.

    \item \textbf{Structural Correction.}
    When a gap is detected, the system inserts intermediate nodes with appropriate dimensions to fill the skipped rank.
    When a dimension is finer-grained than its rank position suggests, it is marked as a \emph{sub-layer}, preserving the original classification without physical restructuring.
    Redundant single-child intermediate nodes at the same rank are merged.
\end{enumerate}

Level alignment operates entirely as a post-processing step, requiring no changes to the core construction loop.
This design keeps the construction phase scalable, since alignment need only run once after the tree is complete, rather than adding per-node overhead.

\section{Evaluation}
\label{sec:eval}

\subsection{Research Questions}
\label{sec:rqs}

We evaluate \aitd{} through three research questions:

\begin{itemize}
  \item \textbf{RQ1 (Taxonomy Quality):} How does \aitd{} compare to baselines in taxonomy quality?
  \item \textbf{RQ2 (Downstream Utility):} How useful is the \aitd{} taxonomy for downstream software engineering tasks?
  \item \textbf{RQ3 (Design Decisions):} What is the contribution of each design decision to the overall quality?
\end{itemize}

\subsection{Experimental Setup}
\label{sec:setup}

\subsubsection{Dataset}
\label{sec:dataset}

\aitd{} requires repositories with sufficient textual metadata for meaningful classification.
We start from all 55,083 GitHub repositories with $\geq$1,000 stars
as of February 20, 2026, which we term \emph{community-adopted
projects}, repositories that have attracted significant developer
attention~\cite{borges2018stars}.
While star-based filtering has been critiqued for selection bias in code
analysis studies~\cite{maj2024stars}, our contribution is a taxonomy
construction methodology independent of the specific population; the
threshold defines the ecosystem we organize, not the validity of our method.

\noindent\textbf{Repository Summaries.}
Raw GitHub descriptions are often too brief for reliable classification
(median $\approx$60 characters).
We construct richer representations through a two-stage pipeline.
First, we extract the overview page from each repository's
DeepWiki~\cite{deepwiki} wiki (median 9,135 characters), which provides
AI-generated documentation covering architecture, functionality, and
technical stack.
At crawl time, DeepWiki had indexed 46,786 repositories (84.9\%);
we manually triggered indexing for the remaining 8,297 through DeepWiki's web interface.
After manual indexing, 696 repositories still could not be indexed
(primarily non-code projects such as resource collections), leaving
\textbf{54,387 repositories} with successful overviews.
Second, we compress each overview into a concise summary
(median 982 characters) using Claude Sonnet, retaining key
information about functionality, technical stack, and use cases.
Generating all 54,387 summaries required 54,387 Sonnet calls
($\approx$148M input tokens, $\approx$13M output tokens),
a one-time preprocessing cost shared across all methods.

\noindent\textbf{Two Experimental Runs.}
We use two complementary datasets, each serving distinct research questions:

\begin{itemize}
  \item \textbf{Full 54k} (54,387 repositories):
    Used for RQ2 (downstream tasks) and ecosystem analysis (Section~\ref{sec:insights}).
    Downstream tasks require comprehensive coverage to avoid missing relevant repositories.
  \item \textbf{Stratified 2k} (2,001 repositories):
    Used for RQ1 (quality comparison) and RQ3 (ablation).
    We apply stratified random sampling~\cite{baltes2022sampling} across three
    star-count buckets (1k--5k, 5k--20k, 20k+), drawing 667 repositories per bucket.
    Equal allocation ensures consistent statistical power and
    eliminates popularity bias from top-$N$ selection.
\end{itemize}

We do not substitute subsets of the full 54k taxonomy for the 2k evaluation:
subtrees built within a 54k context differ structurally from taxonomies
built independently on 2k repositories, and all baselines must run on
identical input for a fair comparison.

\subsubsection{Comparison Methods}
\label{sec:baselines}

As the first work on automated taxonomy construction for software repositories,
no directly comparable prior methods exist.
We select baselines representing four distinct paradigms, ensuring each core
design dimension of \aitd{} is challenged by at least one alternative.
Table~\ref{tab:baselines} summarizes all methods.

\begin{table}[t]
  \caption{Comparison methods and their characteristics.}
  \label{tab:baselines}
  \centering
  \small
  \begin{tabular}{@{}llcc@{}}
    \toprule
    \textbf{Method} & \textbf{Paradigm} & \textbf{RQ1} & \textbf{RQ2} \\
    \midrule
    \aitd{} (ours)    & Multi-agent iterative    & \checkmark & \checkmark \\
    CoL-blind         & LLM knowledge only       & \checkmark &            \\
    CoL-topics        & LLM + seed entities      & \checkmark &            \\
    Single-LLM        & One-shot generation      & \checkmark &            \\
    Emb-mpnet          & Embedding clustering     & \checkmark &            \\
    GitHub Topics     & Community folksonomy     &            & \checkmark \\
    Embedding         & Vector similarity        &            & \checkmark \\
    \bottomrule
  \end{tabular}
\end{table}

\noindent\textbf{RQ1 Baselines} (taxonomy quality on stratified 2k):

\textit{CoL-blind}~\cite{zeng2024col} builds a taxonomy layer by layer using pure LLM
knowledge without any repository data as input.
After construction, we apply \aitd{}'s Classifier Agent to assign repositories
to the resulting tree for evaluation.
This baseline tests whether LLM knowledge alone can produce a taxonomy
suitable for classifying real repositories.
We note that CoL was designed for taxonomy construction from seed terms,
not for item classification; we add a classification step to enable
evaluation on our metrics, which may disadvantage CoL relative to joint design-and-classify methods.

\textit{CoL-topics} uses the same Chain-of-Layer algorithm but seeds it with
GitHub Topics from the dataset as initial entities.
This tests whether real-world vocabulary improves CoL's organization.

\textit{Single-LLM} generates a complete taxonomy in a single LLM call,
receiving a stratified sample of repositories and aggregated topic statistics.
It serves as both the simplest baseline (no iteration, no multi-agent separation)
and the most extreme ablation variant.

\textit{Emb-mpnet} embeds repository summaries using
all-mpnet-base-v2~\cite{reimers2019sentence} and applies recursive Ward
hierarchical clustering with a target branching factor of 6--8.
Category names are generated post-hoc by an LLM summarizing cluster contents.
This represents the standard unsupervised approach.

\noindent\textbf{RQ2 Baselines} (downstream utility on full 54k):

We compare three fundamentally different information organization paradigms:
\aitd{} (taxonomy-based), GitHub Topics (tag-based, the current state of
practice), and cosine-similarity retrieval over mpnet embeddings
(similarity-based).
We do not include CoL or Single-LLM in RQ2 because they cannot scale to
54k repositories, and RQ1 already establishes \aitd{}'s quality advantage
over them.
TaxoAdapt~\cite{taxoadapt2025} is excluded as a baseline because it
requires users to predefine a fixed set of dimension names (e.g.,
tasks, methods, datasets) and builds a separate taxonomy per
dimension, a multi-dimensional design that does not apply to our
setting, where each node dynamically selects its own splitting
dimension within a single unified taxonomy.

\noindent\textbf{RQ3 Ablation Variants} (design decisions on stratified 2k):

To isolate the contribution of each design component, we create four
ablation variants, each removing exactly one element:
\textit{no\_data\_driven} removes repository samples from the Designer
Agent's prompt, relying on pure LLM knowledge for dimension design;
\textit{no\_iteration} disables the iterative refinement loop, running
a single Design$\to$Classify pass per node;
\textit{no\_alignment} compares the taxonomy before and after
Level Alignment (build vs.\ aligned);
\textit{Single-LLM} (also an RQ1 baseline) removes multi-agent
separation, iteration, and level-by-level construction simultaneously.

\subsubsection{Implementation Details}
\label{sec:implementation}

\aitd{} uses Claude Opus 4.6 as the \designeragent{} and Claude Sonnet 4.6
as the \classifieragent{}.
All LLM-based baselines use matching model configurations;
Emb-mpnet uses Claude Opus only for post-hoc cluster naming.
Our replication package releases all prompt templates, including the
dimension-design prompts that operationalize the upper-level stability
principle (Section~\ref{sec:designer}), together with the pre-computed
summaries, generated taxonomies, and evaluation logs.
Optional human review (Algorithm~\ref{alg:single-layer}) is disabled in
all experiments for reproducibility; under full automation only 2 of
2,001 repositories
(2k) and 23 of 54,387 (full) stop short of a leaf category, so the
refinement loop---not manual review---drives leaf completion
(disabling refinement instead leaves 296 of 2,001 short).
Enabling human review would route these few residual cases to finer
leaves; given their small number, its effect on the reported metrics is
correspondingly limited.
For the stratified 2k run, \aitd{} consumed $\approx$6.6M tokens
(1,243 calls); the full 54k run consumed $\approx$188M input +
42M output tokens (38,728 calls).
Baselines require 5--40$\times$ fewer tokens (166k--1.2M),
reflecting \aitd{}'s iterative per-node loop (Section~\ref{sec:rq3}).

\noindent\textbf{Cost.}
The stratified 2k run costs only \$43 and is the practical entry point
for end-to-end replication; to make the larger run reproducible without
rerunning it, we release the full 54k taxonomy outputs and all
pre-computed repository summaries.
Constructing the full 54k taxonomy costs $\approx$\$1{,}500 at current
API pricing, plus a one-time \$640 summary cost shared
across all methods.
The total experimental budget (all methods and evaluation) is $\approx$\$3{,}000.

\noindent\textbf{LLM Judge Reliability.}
Since \aitd{} uses Claude for construction and seven of the eight metrics
rely on LLM-as-judge scoring (all but Coverage), we cross-validate all judge-based
results with GPT-5.4 (an independent model family).
The primary judges are Claude Opus 4.6 for LITE, PG, and SC, and
Claude Sonnet 4.6 for NCP and node relevance.
For RQ1, Pearson $r = 0.995$ across 42 method--metric pairs with
identical TQF rankings (parenthesized values in Table~\ref{tab:rq1}).
For RQ2, cross-validation results are shown in
Figure~\ref{fig:alt_jpk}.
We further validate with human annotation on 250 stratified samples
(oversampling disagreement zones); population-weighted
$\kappa$~\cite{landis1977kappa} is 0.808 for alternative discovery
and 0.932 for retrieval (both ``almost perfect'').
For NCP, human--LLM Pearson $r = 0.959$ with fully preserved
method rankings.

\subsubsection{Evaluation Metrics}
\label{sec:metrics}

We combine two established frameworks with a novel composite score.
\textbf{LITE}~\cite{lite2025} provides four LLM-judged metrics
(scored 0--10): Semantic Clarity (SCA), Hierarchical Relationship
Reasoning (HRR), Exclusivity (HRE), and Independence (HRI).
\textbf{TaxoAdapt}~\cite{taxoadapt2025} adds Path Granularity
(PG, binary), Sibling Coherence (SC, 0--1), and Coverage (Cov, 0--1).
\textbf{Node Classification Precision} (NCP, 0--1) evaluates whether
assigned repositories genuinely belong to their leaf categories, filling
a gap in LITE and TaxoAdapt, which assess structure but not
classification accuracy.

\noindent\textbf{Taxonomy Quality F-score (TQF).}
Individual metrics can be misleading (CoL-blind achieves the highest
LITE but the lowest NCP).
TQF balances structural quality and practical applicability:
\begin{equation}
  \text{TQ} = \frac{1}{3}\!\left(\frac{\text{LITE}_{\text{avg}}}{10} + \text{PG} + \text{SC}\right)
\end{equation}
\begin{equation}
  \text{TQF} = \frac{2 \cdot \text{TQ} \cdot \text{NCP}}{\text{TQ} + \text{NCP}} \times \text{Cov}
\end{equation}
TQF takes the harmonic mean of TQ and NCP, then applies Coverage
as a multiplicative completeness penalty.
Rankings are robust: replacing harmonic mean with arithmetic or
geometric mean, varying TQ--NCP weights from 0.3:0.7 to 0.7:0.3,
or removing Coverage all preserve \aitd{}'s first-place ranking.
For RQ2, we report Precision at $k$ (P@$k$) and Mean Reciprocal
Rank (MRR), with relevance assessed by an LLM judge.

\subsection{RQ1: Taxonomy Quality}
\label{sec:rq1}

\noindent\textbf{Motivation.}
We assess whether \aitd{}'s multi-agent iterative approach produces
higher-quality taxonomies than methods based on pure LLM knowledge,
tag-seeded generation, embedding clustering, or one-shot generation.

\noindent\textbf{Method.}
We run all five methods on the stratified 2k dataset and evaluate
using the full metric suite (Section~\ref{sec:metrics}), with results
in Table~\ref{tab:rq1}.

\begin{table*}[t]
  \caption{Taxonomy quality comparison on stratified 2k dataset. Best values in \textbf{bold},
  second-best \underline{underlined}. LITE metrics are scored 0--10; TaxoAdapt, NCP, and composite metrics are percentages.
  Parenthesized values in NCP/TQ/TQF are from independent GPT-5.4 re-evaluation
  (Pearson $r = 0.995$ across all 42 method--metric pairs); rankings are identical under both judges.}
  \label{tab:rq1}
  \centering
  \small
  \setlength\aboverulesep{0pt}
  \setlength\belowrulesep{0pt}
  \setlength\extrarowheight{0.4ex}
  \begin{tabular}{@{}l cccc c cc c c cc@{}}
    \toprule
    & \multicolumn{4}{c}{\textbf{LITE (0--10)}} & & \multicolumn{2}{c}{\textbf{TaxoAdapt (\%)}} & & & \multicolumn{2}{c}{\textbf{Composite (\%)}} \\
    \cmidrule(lr){2-5} \cmidrule(lr){7-8} \cmidrule(lr){11-12}
    \textbf{Method} & SCA & HRR & HRE & HRI & LITE$_{\text{avg}}$ & PG & SC & Cov & NCP & TQ & \textbf{TQF} \\
    \midrule
    \aitd{}           & 7.35 & 8.72 & 6.74 & 6.90 & 7.43 & 76.39 & \underline{78.69} & \underline{99.84} & \textbf{91.35}\,{\scriptsize(85.19)} & 76.46\,{\scriptsize(77.46)} & \colorbox{gray!18}{\textbf{83.13}\,{\scriptsize(81.01)}} \\
    CoL-blind         & \textbf{7.88} & \textbf{9.40} & \textbf{7.49} & \textbf{7.08} & \textbf{7.96} & \underline{87.10} & \textbf{90.37} & 94.80 & 43.10\,{\scriptsize(34.33)} & \textbf{85.72}\,{\scriptsize(85.60)} & \colorbox{gray!18}{54.40\,{\scriptsize(46.45)}} \\
    CoL-topics        & 7.34 & 7.94 & 5.59 & 4.77 & 6.41 & 58.10 & 42.30 & \textbf{100.00} & \underline{88.90}\,{\scriptsize(78.40)} & 54.80\,{\scriptsize(61.83)} & \colorbox{gray!18}{\underline{67.80}\,{\scriptsize(69.10)}} \\
    Single-LLM        & \underline{7.87} & \underline{9.08} & \underline{6.75} & \textbf{7.08} & \underline{7.70} & \textbf{91.90} & 64.80 & 87.70 & 73.91\,{\scriptsize(63.89)} & \underline{77.90}\,{\scriptsize(79.03)} & \colorbox{gray!18}{66.54\,{\scriptsize(61.99)}} \\
    Emb-mpnet         & 7.23 & 7.28 & 5.16 & 5.87 & 6.38 & 30.10 & 39.20 & \textbf{100.00} & 76.50\,{\scriptsize(67.17)} & 44.40\,{\scriptsize(55.03)} & \colorbox{gray!18}{56.20\,{\scriptsize(60.50)}} \\
    \bottomrule
  \end{tabular}
\end{table*}

\noindent\textbf{Results.}
\aitd{} achieves the highest TQF (83.13\%), outperforming the
second-best method (CoL-topics, 67.80\%) by 15.33 percentage points.
This advantage stems from \aitd{}'s unique balance between structural
quality and practical applicability: it is the only method with both
TQ $>$ 75\% and NCP $>$ 90\%.

\textit{Structural quality alone is misleading.}
CoL-blind achieves the highest LITE (7.96) and TQ (85.72\%), yet
ranks last in TQF (54.40\%) because its categories, built
from pure LLM knowledge, fail to accommodate real repositories
(NCP = 43.10\%).
This validates TQF: evaluating structure alone would rank CoL-blind first.

\textit{Iterative refinement bridges the gap.}
Single-LLM sees repository data yet achieves only moderate NCP
(73.91\%); \aitd{}'s refinement loop raises NCP to 91.35\%,
demonstrating that seeing data once is insufficient; the closed-loop
between design and classification is essential.
CoL-topics achieves competitive NCP (88.90\%) via GitHub Topics as an indirect data signal, but at the cost of low TQ (54.80\%).

\begin{figure}[t]
  \centering
  \includegraphics[width=0.85\columnwidth]{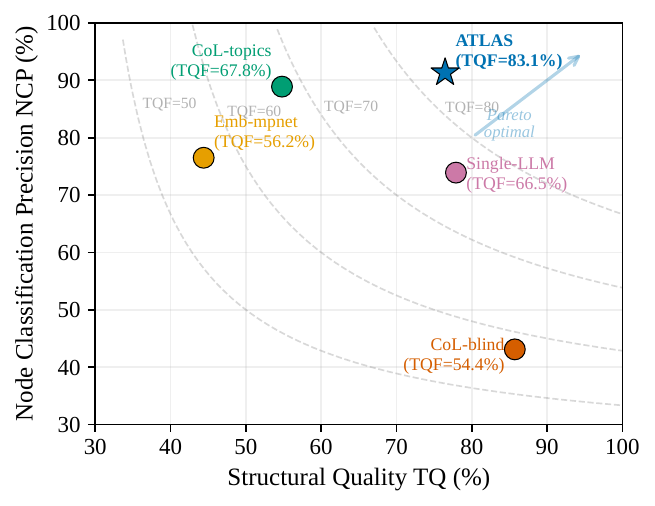}
  \caption{TQ vs.\ NCP tradeoff. Dashed curves are iso-TQF contours.
    \aitd{} (star) is the only method in the Pareto-optimal region
    (high structure \emph{and} high applicability).}
  \label{fig:tq-ncp}
  \Description{Scatter plot of five methods on TQ (x-axis) vs NCP
    (y-axis), with iso-TQF curves. ATLAS is in the upper-right,
    CoL-blind in the lower-right.}
\end{figure}

\textit{Embedding clustering lacks interpretability.}
Emb-mpnet achieves the lowest TQ (44.40\%), confirming that similarity-based clustering cannot produce meaningful hierarchical distinctions.
Figure~\ref{fig:tq-ncp} visualizes this tradeoff: \aitd{} occupies the Pareto-optimal corner, while CoL-blind remains in the high-TQ, low-NCP zone.

\finding{1}{\aitd{} achieves TQF = 83.13\%, outperforming all baselines
by $\geq$15 percentage points. It is the only method that balances
high structural quality (TQ = 76.46\%) with high practical applicability
(NCP = 91.35\%). Evaluating structure alone is misleading: the
highest-LITE method (CoL-blind) has the lowest TQF.}

\subsection{RQ2: Downstream Utility}
\label{sec:rq2}

\noindent\textbf{Motivation.}
Beyond intrinsic quality, a taxonomy must demonstrate practical value.
These two tasks instantiate the primary-stakeholder needs of
Section~\ref{sec:scope}: locating functionally equivalent projects and
navigating to a category of interest.
We evaluate \aitd{}'s 54k taxonomy on two downstream tasks that represent
complementary search scenarios: \emph{alternative discovery}
(given a known repository, find functionally similar projects) and
\emph{repository retrieval} (given a topic query, find relevant repositories).
These tasks compare three fundamentally different information organization
paradigms: taxonomy-based (\aitd{}), tag-based (GitHub Topics), and
similarity-based (embedding cosine distance).

\subsubsection{Alternative Discovery}
\label{sec:alt-discovery}

\noindent\textbf{Setup.}
Given a seed repository, each method returns a ranked list of alternatives.
\aitd{} retrieves repositories from the same leaf category, ranked
primarily by tree distance (closer branches first) and secondarily by
embedding similarity within the same distance;
GitHub Topics ranks by Jaccard similarity over shared topics;
Embedding ranks by cosine similarity of mpnet vectors.
We use two complementary seed sets.
Our primary, human-curated ground truth is
\emph{awesome-oss-alternatives}~\cite{awesomeoss} (154 seeds from
RunaCapital's curated open-source alternative directory, where
categories group projects by domain rather than strict functional
equivalence).
To extend coverage to niche or recently created projects that curated
lists may omit, we add an \emph{LLM-generated} set (200 seeds with
candidate alternatives produced by Claude Opus); because these seeds
share a model family with the \designeragent{} and the judge, we treat
them as a secondary coverage extension rather than primary evidence.
Following standard pooling methodology~\cite{voorhees2005trec},
for each seed we merge the top-10 results from all methods,
deduplicate, and assess relevance via an LLM judge (Claude Opus) in a blind setting with repository summaries only.

\noindent\textbf{Results.}
\aitd{} achieves the highest P@1 and MRR across both ground-truth
sets (Figure~\ref{fig:alt_jpk}a--b;
\ifdefined\ATLASWithoutAppendix
full numerical results are available in the replication package),
\else
full numerical results are in Appendix Table~\ref{tab:alt}),
\fi
\emph{including outperforming the
human-curated lists themselves}.
\aitd{}'s P@1 reaches 85.71\% (awesome-oss) and 79.50\% (LLM-generated),
with MRR of 88.17\% and 83.51\%,
exceeding even the curated baselines (P@1 = 62.34\% / 49.00\%) by over 23
and 30 percentage points respectively.
This occurs because taxonomy leaf categories group repositories by
precise functional equivalence (e.g., self-hosted file storage platforms),
while curated lists often mix projects of different granularity.

GitHub Topics performs worst (P@1 = 55.56\% / 37.43\%), as tag-based
matching lacks the structural precision to distinguish true functional
alternatives from topically related but non-substitutable projects.
Embedding retrieval is competitive (P@1 = 61.69\% / 59.50\%) but
consistently below \aitd{}, confirming that semantic similarity alone
is insufficient for alternative discovery, since projects can be similar in
description but serve different purposes.

\noindent\textbf{Case study.}
For \texttt{kubernetes} (container orchestration), \aitd{} achieves
P@5 = 80\% by retrieving true orchestration alternatives
(\texttt{mesos}, \texttt{dcos}, \texttt{service-fabric},
\texttt{lastbackend}). All other methods score 0\%: Topics returns
K8s ecosystem tools, and Embedding retrieves non-substitutable
extensions.
Conversely, for \texttt{hoppscotch} (API testing), Topics achieves
P@5 = 100\% via direct tag matching, while \aitd{} scores 20\%,
confirming that tags remain effective when precise labels exist.

\subsubsection{Repository Retrieval}
\label{sec:retrieval}

\noindent\textbf{Setup.}
We construct 200 compound queries from GitHub topic pairs
(e.g., ``cli + docker''), stratified by intersection size.
\aitd{} uses beam search (beam width=10, Claude Sonnet) to traverse
category nodes, then ranks repositories within matched leaf
categories by embedding similarity.
GitHub Topics returns repositories tagged with both topics, ranked
by stars; Embedding ranks by cosine similarity to the query text.

\begin{figure*}[t]
  \centering
  \includegraphics[width=0.95\textwidth]{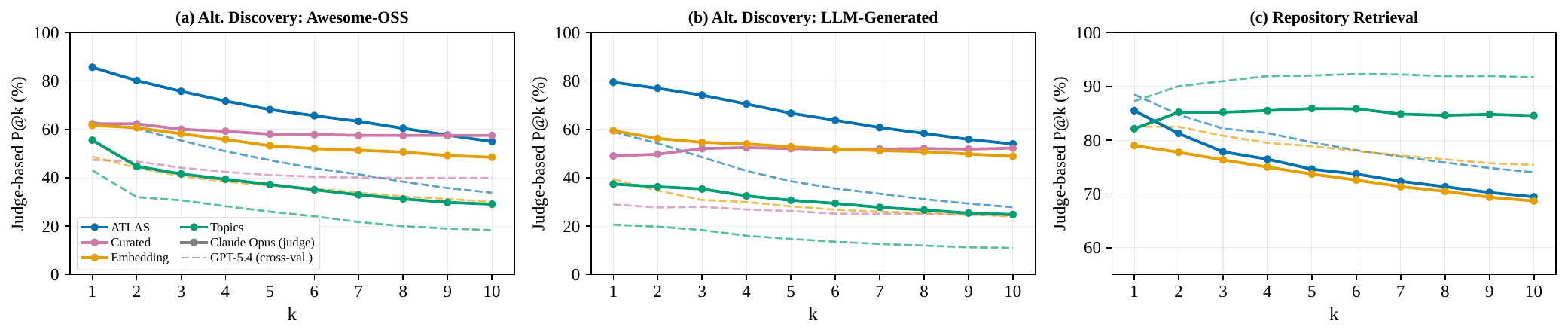}
  \caption{Downstream task P@$k$ ($k = 1$--$10$) under two
    independent LLM judges. Solid lines: Claude Opus (primary);
    dashed lines: GPT-5.4 (cross-validation).
    (a--b)~Alternative discovery: GPT-5.4 applies a stricter
    definition, uniformly lowering precision, but \aitd{} ranks
    first at every $k$ under both judges.
    (c)~Retrieval: queries are constructed from topic pairs,
    giving Topics an inherent advantage at higher $k$; \aitd{}
    nonetheless achieves the highest P@1 under both judges.}
  \label{fig:alt_jpk}
  \Description{Three line charts. Left and center: alternative
    discovery P@k for two seed sets, ATLAS highest under both judges.
    Right: retrieval P@k, Topics overtakes ATLAS at higher k, both
    judges agree on this pattern.}
\end{figure*}

\noindent\textbf{Results.}
Figure~\ref{fig:alt_jpk}c reveals a complementary pattern
\ifdefined\ATLASWithoutAppendix
(full results are available in the replication package).
\else
(full results are in Appendix Table~\ref{tab:retrieval}).
\fi
\aitd{} achieves the highest P@1 (85.50\%, MRR = 88.31\%),
while GitHub Topics achieves higher P@5+ (85.88\% vs.\ 74.61\%), partly
expected since queries are constructed from topic pairs, giving
Topics an inherent exact-match advantage.
The P@$k$ curves show this complementarity: \aitd{}'s precision drops
beyond leaf boundaries, while Topics maintains near-flat
precision.
Taxonomy-based organization provides \emph{depth} (precise top
results), while tag-based organization provides \emph{breadth}
(stable recall).
For example, for ``docker + machine-learning,'' \aitd{} achieves
P@5 = 100\% via the ML deployment category, while
Topics (P@5 = 60\%) includes monitoring and resource-list projects.

\noindent\textbf{Judge cross-validation.}
GPT-5.4 cross-validation (Figure~\ref{fig:alt_jpk}, dashed lines) confirms all method rankings are preserved under both judges.
The two judges calibrate in \emph{opposite} directions: GPT-5.4 is
stricter for alternative discovery ($\kappa = 0.51$, reflecting
inherent task ambiguity) but more lenient for retrieval
($\kappa = 0.78$), ruling out shared systematic bias.
\aitd{}'s positive judgments are the most robust across judges:
51.8\% of Claude-positive \aitd{} pairs survive GPT-5.4's stricter
threshold, versus 36--49\% for other methods.

\finding{2}{\aitd{}'s taxonomy enables superior alternative discovery,
with P@1 = 85.71\% outperforming even human-curated lists (62.34\%).
Taxonomy-based retrieval exhibits a characteristic precision profile:
high P@1 from precise within-category matching, with a cliff-like
drop at higher $k$ when retrieval crosses granularity boundaries.
This makes taxonomies and tags fundamentally complementary: taxonomies
excel at precision-critical tasks (alternative discovery), while tags
provide stable recall for broad exploration (retrieval).}

\subsection{RQ3: Design Decisions}
\label{sec:rq3}

\noindent\textbf{Motivation.}
We isolate the contribution of each design component by removing
one element at a time, holding all else constant.

\noindent\textbf{Method.}
Table~\ref{tab:ablation} reports four ablation variants on the
stratified 2k dataset.
\textit{no\_data\_driven} removes repository samples from the
Designer Agent's prompt;
\textit{no\_iteration} disables the refinement loop;
\textit{no\_alignment} compares the taxonomy before Level Alignment
(build) with the aligned version;
\textit{Single-LLM} (from RQ1) removes multi-agent separation, iteration, and level-by-level construction at once.

\begin{table}[t]
  \caption{Ablation study on stratified 2k. $\Delta$TQF shows the
  change from full \aitd{}.}
  \label{tab:ablation}
  \centering
  \small
  \setlength\aboverulesep{0pt}
  \setlength\belowrulesep{0pt}
  \setlength\extrarowheight{0.4ex}
  \begin{tabular}{@{}l ccc cc@{}}
    \toprule
    \textbf{Variant} & LITE$_{\text{avg}}$ & TQ & NCP & \colorbox{gray!18}{\textbf{TQF}} & $\Delta$ \\
    \midrule
    \aitd{} (full)      & 7.43 & 76.46 & \textbf{91.35} & \colorbox{gray!18}{\textbf{83.13}} & --- \\
    \quad $-$data-driven   & 7.43 & 73.59 & 78.78 & \colorbox{gray!18}{75.98} & $-$7.15 \\
    \quad $-$iteration     & 7.38 & 75.21 & 90.21 & \colorbox{gray!18}{81.39} & $-$1.74 \\
    \quad $-$alignment     & 7.44 & 75.79 & 91.20 & \colorbox{gray!18}{82.65} & $-$0.48 \\
    \midrule
    Single-LLM          & 7.70 & 77.90 & 73.91 & \colorbox{gray!18}{66.54} & $-$16.59 \\
    \bottomrule
  \end{tabular}
\end{table}

\noindent\textbf{Results.}

\textit{Data-driven design is critical for practical applicability.}
Removing repository samples (\textit{no\_data\_driven}) causes the
largest single-component drop ($-$7.15 TQF), driven by an NCP
collapse from 91.35\% to 78.78\% while LITE remains unchanged (7.43).
Data-driven design is essential not for structural quality but for
ensuring the taxonomy accommodates real-world projects.

\textit{Iterative refinement improves structure.}
Disabling the refinement loop (\textit{no\_iteration}) has a moderate effect ($-$1.74 TQF), primarily through reduced TQ (76.46 $\to$ 75.21) while NCP stays high (90.21\%).

\textit{Quality at scale.}
On the full 54k dataset (spot-checking 300 samples per metric),
core metrics remain stable: LITE$_{\text{avg}}$ = 7.18
($\Delta = -0.25$) and NCP = 87.84\% ($\Delta = -3.5$~pp).
PG shows a larger decline to 45.00\% ($\Delta = -31.4$~pp),
because PG assigns a binary score to \emph{entire} root-to-leaf
paths; the 54k tree's longer paths (mean 6.1 vs.\ 4.8 nodes)
amplify this all-or-nothing penalty.
The approximate 54k TQF is 73.0\%, the gap driven by PG, not classification accuracy.

\textit{Level Alignment is structural, not metric-driven.}
Alignment contributes only $-$0.48 TQF on the 2k dataset, but its
value is structural: it establishes a unified rank system ensuring
cross-branch granularity consistency, a prerequisite for the
ecosystem analysis in Section~\ref{sec:insights}.
On the full 54k taxonomy (8 semantic ranks, max depth 10, 36,986
nodes), alignment adjusted 14,580 node depths.

\textit{The full architecture matters most.}
Single-LLM drops TQF by 16.59 points to 66.54\%.
Its TQ (77.90\%) exceeds \aitd{}'s (76.46\%), confirming that a
single LLM can produce structurally sound taxonomies, but without
iterative refinement and data-driven design, it cannot reliably
classify real repositories (NCP = 73.91\%).

\textit{\designeragent{} sample size.}
The \designeragent{} samples a fixed number of repositories per node
(default 80, Section~\ref{sec:designer}).
Table~\ref{tab:samplesize} sweeps this budget over 40, 80, and 160 on
the stratified 2k dataset.
TQF improves monotonically (81.97 $\to$ 83.13 $\to$ 83.65) but with
clear diminishing returns: $+1.16$ points from 40 to 80, then only
$+0.52$ from 80 to 160 at double the cost.
The two structural components move in opposite directions: Path
Granularity rises steadily (73.82 $\to$ 82.63) as larger samples let
the \designeragent{} choose finer dimensions, while Sibling Coherence peaks at
$n = 80$ (78.69) and declines at $n = 160$ (74.03), indicating that
the largest budget begins to over-split categories.
Read against the \textit{no\_data\_driven} ablation---the zero-sample
endpoint ($-7.15$ TQF)---$n = 80$ sits at the quality/cost knee: it
captures most achievable quality and the best sibling
consistency without over-splitting or doubled cost of $n = 160$.

\begin{table}[t]
  \caption{\designeragent{} sample-size sweep on stratified 2k.
  LITE is scored 0--10; other columns are percentages.}
  \label{tab:samplesize}
  \centering
  \small
  \setlength\aboverulesep{0pt}
  \setlength\belowrulesep{0pt}
  \setlength\extrarowheight{0.4ex}
  \begin{tabular}{@{}l cccc c c@{}}
    \toprule
    \textbf{Sample $n$} & LITE$_{\text{avg}}$ & PG & SC & Cov & NCP & \colorbox{gray!18}{\textbf{TQF}} \\
    \midrule
    40                & 7.24 & 73.82 & 74.44 & \textbf{100.00} & \textbf{92.55} & \colorbox{gray!18}{81.97} \\
    80 (default)      & \textbf{7.43} & 76.39 & \textbf{78.69} & 99.84 & 91.35 & \colorbox{gray!18}{83.13} \\
    160               & 7.35 & \textbf{82.63} & 74.03 & \textbf{100.00} & 91.95 & \colorbox{gray!18}{\textbf{83.65}} \\
    \bottomrule
  \end{tabular}
\end{table}

\textit{Model generalizability.}
Table~\ref{tab:model-config} reports results with three alternative
model configurations on the stratified 2k dataset.

\begin{table}[t]
  \caption{Model configuration comparison on stratified 2k.}
  \label{tab:model-config}
  \centering
  \small
  \setlength\aboverulesep{0pt}
  \setlength\belowrulesep{0pt}
  \setlength\extrarowheight{0.4ex}
  \begin{tabular}{@{}l cc cc r@{}}
    \toprule
    \textbf{Configuration} & LITE$_{\text{avg}}$ & TQ & NCP & \colorbox{gray!18}{\textbf{TQF}} & Tokens \\
    \midrule
    Opus + Sonnet (default)   & \textbf{7.43} & 76.46 & \textbf{91.35} & \colorbox{gray!18}{\underline{83.13}} & 6.6M \\
    Opus + Opus               & 7.33 & \textbf{77.09} & \underline{90.95} & \colorbox{gray!18}{\textbf{83.45}} & 6.1M \\
    GPT-5.4 + GPT-5.4        & 7.04 & 73.44 & 90.27 & \colorbox{gray!18}{80.76} & 7.5M \\
    MiniMax-M2.5 + M2.5      & 6.76 & 64.77 & 83.48 & \colorbox{gray!18}{72.89} & 8.6M \\
    \bottomrule
  \end{tabular}
\end{table}

All configurations achieve TQF $> 72\%$, confirming that the
architecture, not the specific model, drives quality.
Opus+Sonnet and Opus+Opus perform nearly identically
(83.13\% vs.\ 83.45\%), validating asymmetric pairing.
Dimension design quality varies more across model families than
classification accuracy (MiniMax's largest gap is in PG).

\finding{3}{Data-driven design is the most impactful single component
($\Delta$TQF = $-$7.15), primarily through practical applicability
(NCP) rather than structural quality.
Iterative refinement contributes to structural polish
($\Delta$TQF = $-$1.74).
The combined architecture (multi-agent + iteration + data-driven)
yields a 16.59-point TQF advantage over one-shot generation.
The architecture generalizes across model families (TQF $> 72\%$
for all tested configurations), with asymmetric pairing
(strong \designeragent{} + light \classifieragent{}) matching symmetric
configurations at lower cost.
The \designeragent{}'s sample budget is well-behaved:
quality rises monotonically but saturates, leaving
$n = 80$ at the quality/cost knee.}

\ifdefined\ATLASCameraReady
\input{sections/discussion_cr}
\input{sections/related_cr}
\input{sections/conclusion_cr}
\else
\section{Discussion}
\label{sec:discussion}

\subsection{Ecosystem Insights}
\label{sec:insights}

Beyond evaluating \aitd{} as a method, the 54{,}387-repository taxonomy
reveals structural patterns in the open-source ecosystem.
We highlight three observations that depend on the taxonomy's hierarchical
type-based organization---specifically the L1 \emph{artifact type}
dimension (Library, Application, Platform, etc.), which neither
GitHub Topics nor embedding clustering produces.

\observation{1}{The ecosystem follows a long-tail distribution
with domain-specific concentration patterns.}

\noindent At the L2 level (141 categories), the top 31\% of categories contain
80\% of all repositories (Gini = 0.647), while the bottom 50\% hold
only 6\%.
Concentration varies markedly across domains:
Plugin/Extension is highly fragmented (27 L2 subcategories,
HHI = 0.091), reflecting diverse extension ecosystems across
editors, browsers, and frameworks, while Framework/Boilerplate
is winner-take-all (16 subcategories, but top 3 cover 62\%,
HHI = 0.159).
This competitive-landscape difference is only visible through
hierarchical categorization that distinguishes artifact types.

\observation{2}{The ecosystem is shifting from libraries to
applications and platforms.}

\noindent Fig.~\ref{fig:ecosystem}(a) shows L1 domain proportions by
repository creation year.
Library/SDK declined from 45\% (2008) to 13\% (2025)
(Mann-Kendall $\tau = -0.935$, $p < 0.001$,
Sen's slope $= -2.3$~pp/year), while Standalone Application
rose from 9\% to 28\% ($\tau = +0.948$, $p < 0.001$) and
Platform/Service grew from 5\% to 16\%
($\tau = +0.647$, $p < 0.001$).
This reflects ecosystem maturation: as foundational libraries stabilize, effort shifts toward end-user products and integrated platforms.

\observation{3}{AI/ML now dominates newly adopted projects.}

\noindent Fig.~\ref{fig:ecosystem}(b) traces the proportion of AI/ML-related
repositories among newly community-adopted projects (those reaching
$\geq$1{,}000 stars).
The share grew steadily from 12\% (2012) to 29\% (2022), then
jumped 19 percentage points to 48\% in 2023---coinciding with
ChatGPT's release---and reached 61\% by 2025
($\tau = +0.956$, $p < 0.001$, Sen's slope $= +2.5$~pp/year).
More than half of recently community-adopted open-source projects
are now AI/ML-related.

\begin{figure}[t]
  \centering
  \includegraphics[width=\columnwidth]{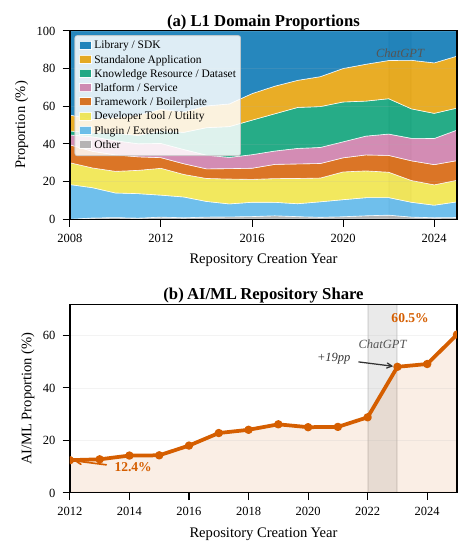}
  \caption{Ecosystem structure and evolution based on the \aitd{}
    taxonomy of 54{,}387 repositories ($\geq$1{,}000 stars).
    (a)~L1 domain proportions by creation year: Library/SDK share
    declined from 45\% to 13\%, while Standalone Application rose
    from 9\% to 28\%.
    (b)~AI/ML repository share surged from 12\% (2012) to 61\%
    (2025), with a 19-percentage-point jump in 2023 after
    ChatGPT's release.}
  \label{fig:ecosystem}
  \Description{Top panel: stacked area chart showing L1 category
    proportions from 2008 to 2025, with Library/SDK declining and
    Application/Platform growing. Bottom panel: line chart showing
    AI/ML proportion rising from 12\% to 61\%, with a sharp
    inflection in 2023.}
\end{figure}

All three observations require the hierarchical, type-based structure that \aitd{} produces; none would emerge from flat tags or embedding clusters alone.

\subsection{Threats to Validity}
\label{sec:threats-detail}

\noindent\textbf{Internal validity.}
LLM non-determinism may produce different taxonomies across runs;
self-corrective refinement and Level Alignment mitigate this.
Across three independent runs on the stratified 2k dataset,
TQF ranges from 83.1\% to 83.7\% (std = 0.3\%).
LLM judge bias is mitigated through cross-validation with GPT-5.4
(Pearson $r = 0.995$, identical rankings; Table~\ref{tab:rq1})
and human annotation ($\kappa \geq 0.808$;
Section~\ref{sec:implementation}).

\noindent\textbf{External validity.}
Our evaluation uses GitHub repositories with $\geq$1,000 stars,
which may not generalize to other ecosystems or below-threshold
repositories.
However, \aitd{}'s framework is domain-agnostic and could be applied
to other domains.
The summary pipeline depends on DeepWiki, a third-party service whose
long-term availability we cannot guarantee.
To bound this dependence we apply DeepWiki summaries \emph{uniformly}
to \aitd{} and all baselines (Section~\ref{sec:dataset}), so it is not
an \aitd{}-specific advantage, and we release every pre-computed
summary so that results reproduce without re-crawling.
README- or documentation-derived summaries are compatible with the
architecture but lie outside our evaluated scope.

\noindent\textbf{Single-label scope.}
\aitd{} assigns each repository a single \emph{dominant functional
role}, and multi-label classification is outside its current scope.
The evaluated tasks (alternative discovery and retrieval) are driven
by a repository's primary function, so they chiefly exercise the
dominant-role taxonomy rather than secondary facets.
Repositories that genuinely span multiple strong roles could benefit
from secondary facets in tasks such as faceted exploration; the
\classifieragent{}'s per-repository confidence scores
(Section~\ref{sec:classifier}) offer a natural hook for flagging such
cases, though a principled multi-label protocol with multiple
annotators is beyond our scope.
This boundary is consistent with the functional, single-taxonomy scope
defined in Section~\ref{sec:scope}.

\noindent\textbf{Construct validity.}
TQF weights TQ and NCP equally via harmonic mean; alternative
weightings could yield different rankings, though we show TQF
resolves a concrete anomaly (CoL-blind: first in structure, last
in applicability).
Downstream evaluation uses LLM-as-judge precision; cross-model
validation confirms preserved rankings
(Sections~\ref{sec:alt-discovery}--\ref{sec:retrieval}).
The model configuration experiment (Table~\ref{tab:model-config}) shows cross-family generalizability, though weaker models (MiniMax) suggest a floor effect on design quality.

\section{Related Work}
\label{sec:related}

\noindent\textbf{Taxonomy Construction.}
Classical taxonomy construction relies on lexical patterns
(Hearst patterns~\cite{hearst1992automatic}), distributional
similarity, or graph-based methods to extract hypernymy relations
from text corpora.
Corpus-driven methods such as TaxoGen~\cite{zhang2018taxogen}
use term embeddings and hierarchical clustering to build topical
taxonomies, while TaxoExpan~\cite{shen2020taxoexpan} expands
existing taxonomies with new concepts via position-enhanced
graph neural networks.
These methods require large text corpora with explicit term
co-occurrence and do not handle the heterogeneous, multi-faceted
nature of software repositories.
Recent work has leveraged LLMs for taxonomy construction.
Chain-of-Layer~\cite{zeng2024col} iteratively prompts an LLM to
build a taxonomy layer by layer, using an ensemble-based ranking
filter to reduce hallucinated relations.
TaxoAdapt~\cite{taxoadapt2025} extends this with density-driven
expansion to adapt multidimensional taxonomies to evolving corpora.
Sas and Capiluppi~\cite{sas2024bottomup} evaluate hybrid approaches
combining Wikipedia, ontologies, and LLMs for software application
domain taxonomy.
Closest to our setting, Nakashima et al.~\cite{nakashima2025satd}
apply an LLM to construct a self-admitted-technical-debt taxonomy and
conclude that \emph{full automation remains an open challenge}---the
gap \aitd{} directly targets.
The two efforts differ on four axes.
\emph{Object:} their pipeline classifies short technical-debt
code-comment snippets, whereas \aitd{} classifies entire repositories
from multi-source documentation.
\emph{Output:} they produce a two-level tree, whereas \aitd{} produces
a multi-rank taxonomy with MECE validation and explicit level
alignment.
\emph{Method:} their explanation-driven, two-phase pipeline generates
and merges categories batch by batch, whereas \aitd{} adds
\designeragent{}/\classifieragent{} separation, a null-rate-driven
refinement loop under minimum intervention, and post-hoc rank discovery.
\emph{Scale:} 448 comments versus our stratified 2k benchmark and full
54k corpus.
These methods target general-purpose or scientific taxonomies, or
finer-grained debt categories;
\aitd{} is, to our knowledge, the first to address software
\emph{repository} taxonomy construction, introducing self-corrective
refinement and data-driven dimension design to handle the
heterogeneity of software artifacts.
The broader field of ontology engineering shares the goal of
hierarchical knowledge organization with explicit
definitions~\cite{cimiano2006ontology}, but typically relies on
knowledge-engineering-driven construction rather than automated,
data-driven methods.
For taxonomy evaluation, we adopt the LITE
framework~\cite{lite2025}, which uses LLM judges with
cross-validation, and TaxoAdapt's granularity and coherence
metrics~\cite{taxoadapt2025}.

\noindent\textbf{Multi-Agent LLM Systems for SE.}
Multi-agent frameworks assign specialized roles to LLM instances
for collaborative problem-solving.
MetaGPT~\cite{hong2024metagpt} encodes software development
SOPs into multi-agent pipelines with distinct PM, architect, and
engineer roles.
ChatDev~\cite{qian2024chatdev} models the software development
lifecycle as agent communication through design, coding, and testing
phases.
In software engineering specifically,
SWE-agent~\cite{yang2024sweagent} and
Agentless~\cite{xia2025agentless} address automated issue resolution,
while MAGIS~\cite{tao2024magis} uses a four-agent framework for
GitHub issue fixing.
\aitd{} differs from these systems in both task and design:
it targets taxonomy construction (a knowledge organization task)
rather than code generation, and its two-agent architecture
(\designeragent{} + \classifieragent{}) reflects a cognitive separation between
schema design and item classification rather than a software
development workflow.

\noindent\textbf{Software Repository Classification.}
GitHub Topics~\cite{izadi2021topic} is the de facto folksonomy
for repository classification but suffers from inconsistency,
incompleteness, and flatness (Section~\ref{sec:bg-taxonomy}).
HiGitClass~\cite{zhang2019higitclass} uses heterogeneous
information network embedding for weakly-supervised hierarchical
classification of GitHub repositories, but relies on predefined
category hierarchies rather than constructing them.
Sas and Capiluppi~\cite{sas2022antipatterns} identify seven
antipatterns in software classification taxonomies and propose
construction guidelines.
GitRanking~\cite{sas2023gitranking} ranks GitHub topics into
301 application domains using active sampling with Bayesian
inference.
A recent systematic mapping study~\cite{balla2025sms} surveys
43 approaches to automatic repository classification, confirming
that most rely on predefined label sets.
These approaches either organize domain \emph{labels} into
hierarchies or classify repositories into \emph{predefined}
categories.
\aitd{} fills the gap between these two tasks: it automatically
constructs a deep, hierarchical taxonomy \emph{and} classifies
54k+ repositories into it end-to-end, through iterative
LLM-driven dimension design.

\section{Conclusion}
\label{sec:conclusion}

We presented \aitd{}, the first framework for automated end-to-end
taxonomy construction and classification of software repositories
at scale.
By separating dimension design from classification through a
multi-agent architecture and connecting them via a self-corrective
refinement loop, \aitd{} produces taxonomies that are both
structurally sound and practically applicable---a balance that
no single-pass or knowledge-only approach achieves.
Evaluation yields a TQF of 83.13\% on a stratified 2k benchmark
(+15~pp over the best baseline; an approximate 73.0\% on the full 54k
corpus, reflecting Path Granularity's all-or-nothing scoring on longer
paths rather than a classification deficit), alternative discovery that
surpasses human-curated lists, and ecosystem insights that reveal the
structural shift from libraries to AI/ML applications.

\aitd{} organizes repositories by their dominant functional role
(Section~\ref{sec:scope}); supporting secondary facets for multi-label,
faceted exploration is a natural extension, alongside cross-domain
application and incremental taxonomy updates.

\begin{acks}
This work is supported in part by the National Key Research and Development
Program of China (No.~2023YFB3307202); and by the National Research Foundation Singapore and
the Cyber Security Agency under the National Cybersecurity R\&D Programme
(NCRP25-P04-TAICeN). This research is also part of the IN-CYPHER Programme
and is supported by the National Research Foundation, Prime Minister's
Office, Singapore, under its Campus for Research Excellence and Technological
Enterprise (CREATE) Programme. Any opinions, findings and conclusions, or
recommendations expressed in these materials are those of the author(s) and
do not reflect the views of the National Research Foundation, Singapore,
Cyber Security Agency of Singapore, Singapore. This work is also supported by
the AI Innovation and Application Joint Research Institute, Shenzhen
Technology University (Grant No.~20261064010117).
\end{acks}

\section{Data-Availability Statement}
\label{sec:data-availability}

All code, data, and replication artifacts are available in the
\aitd{} replication package on Zenodo~\cite{lu2026atlasartifact}.

\fi

\ifdefined\ATLASCameraReady
\else
\clearpage
\fi
\balance
\bibliographystyle{ACM-Reference-Format}
\bibliography{references}

@inproceedings{zhang2018taxogen,
  title={{TaxoGen}: Unsupervised Topic Taxonomy Construction by Adaptive Term Embedding and Clustering},
  author={Zhang, Chao and Tao, Fangbo and Chen, Xiusi and Shen, Jiaming and Jiang, Meng and Sadler, Brian and Vanni, Michelle and Han, Jiawei},
  booktitle={Proceedings of the 24th ACM SIGKDD International Conference on Knowledge Discovery \& Data Mining},
  pages={2701--2709},
  doi={10.1145/3219819.3220064},
  year={2018}
}

@inproceedings{shen2020taxoexpan,
  title={{TaxoExpan}: Self-supervised Taxonomy Expansion with Position-Enhanced Graph Neural Network},
  author={Shen, Jiaming and Shen, Zhihong and Xiong, Chenyan and Wang, Chi and Wang, Kuansan and Han, Jiawei},
  booktitle={Proceedings of The Web Conference 2020},
  pages={486--497},
  doi={10.1145/3366423.3380132},
  year={2020}
}

@inproceedings{balla2025sms,
  title={Automatic Classification of Software Repositories: A Systematic Mapping Study},
  author={Balla, Stefano and Degueule, Thomas and Robbes, Romain and Falleri, Jean-R{\'e}my and Zacchiroli, Stefano},
  booktitle={Proceedings of the 29th International Conference on Evaluation and Assessment in Software Engineering (EASE)},
  doi={10.1145/3756681.3756958},
  year={2025}
}

@inproceedings{nakashima2025satd,
  title={How Far Have {LLMs} Come Toward Automated {SATD} Taxonomy Construction?},
  author={Nakashima, Sota and Ishimoto, Yuta and Kondo, Masanari and Xiao, Tao and Kamei, Yasutaka},
  booktitle={Proceedings of the 32nd Asia-Pacific Software Engineering Conference (APSEC), Early Research Achievements (ERA) Track},
  doi={10.1109/APSEC66846.2025.00087},
  year={2025}
}

@book{cimiano2006ontology,
  title={Ontology Learning and Population from Text: Algorithms, Evaluation and Applications},
  author={Cimiano, Philipp},
  doi={10.1007/978-0-387-39252-3},
  year={2006},
  publisher={Springer}
}

@article{izadi2021topic,
  author = {Izadi, Maliheh and Heydarnoori, Abbas and Gousios, Georgios},
  title = {Topic Recommendation for Software Repositories Using Multi-label Classification Algorithms},
  journal = {Empirical Software Engineering},
  volume = {26},
  number = {5},
  pages = {93},
  doi = {10.1007/s10664-021-09976-2},
  year = {2021},
  publisher = {Springer}
}

@book{mayr1969principles,
  title={Principles of Systematic Zoology},
  author={Mayr, Ernst},
  year={1969},
  publisher={McGraw-Hill},
  address={New York}
}

@book{minto1987pyramid,
  title={The Pyramid Principle: Logic in Writing and Thinking},
  author={Minto, Barbara},
  year={1981},
  publisher={Pitman},
  address={London}
}

@article{borges2018stars,
  title={What's in a GitHub Star? Understanding Repository Starring Practices in a Social Coding Platform},
  author={Borges, Hudson and Valente, Marco Tulio},
  journal={Journal of Systems and Software},
  volume={146},
  pages={112--129},
  doi={10.1016/j.jss.2018.09.016},
  year={2018},
  publisher={Elsevier}
}

@inproceedings{maj2024stars,
  title={The Fault in Our Stars: Designing Reproducible Large-scale Code Analysis Experiments},
  author={Maj, Petr and Muroya, Stefanie and Siek, Konrad and Di Grazia, Luca and Vitek, Jan},
  booktitle={Proceedings of the 38th European Conference on Object-Oriented Programming (ECOOP)},
  doi={10.4230/LIPIcs.ECOOP.2024.27},
  year={2024}
}

@article{baltes2022sampling,
  title={Sampling in Software Engineering Research: A Critical Review and Guidelines},
  author={Baltes, Sebastian and Ralph, Paul},
  journal={Empirical Software Engineering},
  volume={27},
  number={4},
  pages={94},
  doi={10.1007/s10664-021-10072-8},
  year={2022},
  publisher={Springer}
}

@inproceedings{zeng2024col,
  title={Chain-of-Layer: Iteratively Prompting Large Language Models for Taxonomy Induction from Limited Examples},
  author={Zeng, Qingkai and Bai, Yuyang and Tan, Zhaoxuan and Feng, Shangbin and Liang, Zhenwen and Zhang, Zhihan and Jiang, Meng},
  booktitle={Proceedings of the 33rd ACM International Conference on Information and Knowledge Management (CIKM)},
  doi={10.1145/3627673.3679608},
  year={2024}
}

@inproceedings{reimers2019sentence,
  title={Sentence-{BERT}: Sentence Embeddings using Siamese {BERT}-Networks},
  author={Reimers, Nils and Gurevych, Iryna},
  booktitle={Proceedings of the 2019 Conference on Empirical Methods in Natural Language Processing (EMNLP)},
  doi={10.18653/v1/D19-1410},
  year={2019}
}

@article{lite2025,
  title={{LITE}: {LLM}-Impelled Efficient Taxonomy Evaluation},
  author={Zhang, Lin and Gu, Zhouhong and Zheng, Suhang and Wang, Tao and Li, Tianyu and Feng, Hongwei and Xiao, Yanghua},
  journal={arXiv preprint arXiv:2504.01369},
  doi={10.48550/arXiv.2504.01369},
  year={2025}
}

@inproceedings{taxoadapt2025,
  title={{TaxoAdapt}: Aligning {LLM}-Based Multidimensional Taxonomy Construction to Evolving Research Corpora},
  author={Kargupta, Priyanka and Zhang, Nan and Zhang, Yunyi and Zhang, Rui and Mitra, Prasenjit and Han, Jiawei},
  booktitle={Proceedings of the 63rd Annual Meeting of the Association for Computational Linguistics (ACL)},
  pages={29834--29850},
  doi={10.18653/v1/2025.acl-long.1442},
  year={2025}
}

@misc{vander2007folksonomies,
  title={Folksonomy},
  author={Vander Wal, Thomas},
  howpublished={\url{https://vanderwal.net/folksonomy.html}},
  year={2007},
  note={Accessed: March 2026}
}

@misc{awesomeoss,
  title={Awesome Open-Source Alternatives to {SaaS}},
  author={{Runa Capital}},
  howpublished={\url{https://github.com/RunaCapital/awesome-oss-alternatives}},
  year={2024},
  note={Accessed: March 2026}
}

@book{voorhees2005trec,
  title={{TREC}: Experiment and Evaluation in Information Retrieval},
  editor={Voorhees, Ellen M. and Harman, Donna K.},
  year={2005},
  publisher={MIT Press}
}

@inproceedings{hearst1992automatic,
  title={Automatic Acquisition of Hyponyms from Large Text Corpora},
  author={Hearst, Marti A.},
  booktitle={Proceedings of the 14th Conference on Computational Linguistics (COLING)},
  year={1992}
}

@inproceedings{hong2024metagpt,
  title={{MetaGPT}: Meta Programming for a Multi-Agent Collaborative Framework},
  author={Hong, Sirui and Zhuge, Mingchen and Chen, Jonathan and others},
  booktitle={Proceedings of the 12th International Conference on Learning Representations (ICLR)},
  year={2024}
}

@inproceedings{qian2024chatdev,
  title={{ChatDev}: Communicative Agents for Software Development},
  author={Qian, Chen and Liu, Wei and Liu, Hongzhang and others},
  booktitle={Proceedings of the 62nd Annual Meeting of the Association for Computational Linguistics (ACL)},
  doi={10.18653/v1/2024.acl-long.810},
  year={2024}
}

@inproceedings{yang2024sweagent,
  title={{SWE}-agent: Agent-Computer Interfaces Enable Automated Software Engineering},
  author={Yang, John and Jimenez, Carlos E. and Wettig, Alexander and others},
  booktitle={Advances in Neural Information Processing Systems (NeurIPS)},
  year={2024}
}

@article{xia2025agentless,
  title={Demystifying {LLM}-based Software Engineering Agents},
  author={Xia, Chunqiu Steven and Deng, Yinlin and Dunn, Soren and Zhang, Lingming},
  journal={Proceedings of the ACM on Software Engineering},
  volume={2},
  number={FSE},
  doi={10.1145/3715754},
  year={2025},
  publisher={ACM}
}

@inproceedings{tao2024magis,
  title={{MAGIS}: {LLM}-Based Multi-Agent Framework for {GitHub} Issue Resolution},
  author={Tao, Wei and Zhou, Yucheng and Wang, Yanlin and Zhang, Wenqiang and Zhang, Hongyu and Cheng, Yu},
  booktitle={Advances in Neural Information Processing Systems (NeurIPS)},
  year={2024}
}

@article{sas2024bottomup,
  title={Automatic Bottom-Up Taxonomy Construction: A Software Application Domain Study},
  author={Sas, Cezar and Capiluppi, Andrea},
  journal={arXiv preprint arXiv:2409.15881},
  doi={10.48550/arXiv.2409.15881},
  year={2024}
}

@article{sas2022antipatterns,
  title={Antipatterns in Software Classification Taxonomies},
  author={Sas, Cezar and Capiluppi, Andrea},
  journal={Journal of Systems and Software},
  volume={190},
  pages={111343},
  doi={10.1016/j.jss.2022.111343},
  year={2022},
  publisher={Elsevier}
}

@article{sas2023gitranking,
  title={{GitRanking}: A Ranking of {GitHub} Topics for Software Classification using Active Sampling},
  author={Sas, Cezar and Capiluppi, Andrea and Di Sipio, Claudio and Di Rocco, Juri and Di Ruscio, Davide},
  journal={Software: Practice and Experience},
  volume={53},
  number={10},
  pages={1982--2006},
  doi={10.1002/spe.3238},
  year={2023},
  publisher={Wiley}
}

@inproceedings{zhang2019higitclass,
  title={{HiGitClass}: Keyword-Driven Hierarchical Classification of {GitHub} Repositories},
  author={Zhang, Yu and Xu, Frank F. and Li, Sha and Meng, Yu and Wang, Xuan and Li, Qi and Han, Jiawei},
  booktitle={Proceedings of the IEEE International Conference on Data Mining (ICDM)},
  doi={10.1109/ICDM.2019.00098},
  year={2019}
}

@article{landis1977kappa,
  title={The Measurement of Observer Agreement for Categorical Data},
  author={Landis, J. Richard and Koch, Gary G.},
  journal={Biometrics},
  volume={33},
  number={1},
  pages={159--174},
  doi={10.2307/2529310},
  year={1977}
}

@misc{deepwiki,
  title={{DeepWiki}: {AI}-Powered Documentation for Open Source},
  author={{Cognition AI}},
  howpublished={\url{https://deepwiki.com}},
  year={2025},
  note={Accessed: February 2026}
}

@misc{lu2026atlasartifact,
  title={{ATLAS}: Agentic Taxonomy of LArge-Scale Software Ecosystems},
  author={Lu, Junyi and Lyu, Mengyao and Wu, Jiahui and Yu, Lei and Liu, Chengwei and Zhang, Fengjun and Yang, Li and Zuo, Chun and Liu, Yang},
  publisher={Zenodo},
  year={2026},
  doi={10.5281/zenodo.21160280},
  note={Replication package}
}

\ifdefined\ATLASWithoutAppendix
\else
\newpage
\appendix
\section{Downstream Task Detailed Results}
\label{sec:appendix}

\begin{table}[h]
  \caption{Alternative discovery results (judge-based precision, \%).}
  \label{tab:alt}
  \centering
  \small
  \begin{tabular}{@{}l ccccc@{}}
    \toprule
    \textbf{Method} & P@1 & P@3 & P@5 & P@10 & MRR \\
    \midrule
    \multicolumn{6}{@{}l}{\textit{awesome-oss-alternatives} (154 seeds)} \\
    \quad Curated     & \underline{62.34} & \underline{60.06} & \underline{58.05} & \textbf{57.51} & 68.58 \\
    \quad \aitd{}     & \textbf{85.71} & \textbf{75.76} & \textbf{68.18} & \underline{55.00} & \textbf{88.17} \\
    \quad Topics      & 55.56 & 41.34 & 37.01 & 28.90 & 63.65 \\
    \quad Embedding   & 61.69 & 58.23 & 53.25 & 48.51 & \underline{73.18} \\
    \midrule
    \multicolumn{6}{@{}l}{\textit{LLM-generated} (200 seeds)} \\
    \quad Curated     & 49.00 & 52.08 & 52.05 & \underline{52.27} & 65.04 \\
    \quad \aitd{}     & \textbf{79.50} & \textbf{74.17} & \textbf{66.70} & \textbf{54.00} & \textbf{83.51} \\
    \quad Topics      & 37.43 & 31.67 & 27.50 & 22.23 & 45.53 \\
    \quad Embedding   & \underline{59.50} & \underline{54.67} & \underline{52.80} & 48.90 & \underline{69.14} \\
    \bottomrule
  \end{tabular}
\end{table}

\begin{table}[h]
  \caption{Repository retrieval results (judge-based, \%).}
  \label{tab:retrieval}
  \centering
  \small
  \begin{tabular}{@{}l ccccc@{}}
    \toprule
    \textbf{Method} & P@1 & P@3 & P@5 & P@10 & MRR \\
    \midrule
    \aitd{}     & \textbf{85.50} & \underline{77.83} & \underline{74.61} & \underline{69.47} & \underline{88.31} \\
    Topics      & \underline{82.14} & \textbf{83.50} & \textbf{85.88} & \textbf{84.56} & \textbf{90.56} \\
    Embedding   & 79.00 & 76.33 & 73.70 & 68.70 & 85.44 \\
    \bottomrule
  \end{tabular}
\end{table}

\section{Supplementary Details}
\label{sec:supplementary}

\subsection{Implementation Cost Breakdown}
\label{sec:cost-breakdown}

\aitd{} uses Claude Opus as the \designeragent{} and Claude Sonnet
as the \classifieragent{}.
This asymmetric pairing balances capability and cost: the \designeragent{}
requires stronger reasoning for schema design, while the \classifieragent{}
performs a simpler assignment task at higher throughput.

For the stratified 2k run, \aitd{} consumed 1,243 LLM calls
($\approx$6.6M tokens):
769 Opus calls ($\approx$3.0M input tokens) for the \designeragent{}
(design, refinement, validation, and alignment) and
474 Sonnet calls ($\approx$2.6M input tokens) for the
\classifieragent{}---which processes comparable token
volume at substantially lower per-token cost, validating the
asymmetric pairing design.
The full 54k run consumed 38,728 calls
($\approx$188M input tokens, $\approx$42M output tokens),
of which 7,232 were Level Alignment calls.
Baseline methods require substantially fewer tokens:
CoL-blind ($\approx$1.2M), CoL-topics ($\approx$660k),
Single-LLM ($\approx$560k), and Emb-mpnet ($\approx$166k,
embedding-based with LLM only for cluster naming).

At current API pricing (Claude Opus 4.6: \$5/\$25 per MTok
input/output; Sonnet 4.6: \$3/\$15),
constructing the full 54k taxonomy costs approximately \$1{,}500,
plus a one-time \$640 summary generation cost shared across all
methods; the 2k run costs \$43.
Baseline construction ranges from \$1 (Emb-mpnet) to \$8
(CoL-blind).
The full experimental budget including all baselines, model
configurations (GPT-5.4 at \$2.50/\$15; MiniMax-M2.5,
open-source, at \$0.30/\$1.20 via hosted API), evaluation,
and cross-validation totals approximately \$3{,}000.

\aitd{}'s higher token budget (6.6M vs.\ 560k for Single-LLM)
reflects its iterative architecture, not an unfair resource advantage:
the additional tokens are spent on per-node classification
feedback loops that both improve NCP and enable scalability
to 54k repositories---a capability that Single-LLM's one-shot
architecture cannot achieve regardless of token budget.

\subsection{Human Annotation Details}
\label{sec:human-annotation}

We validate LLM judge reliability with human annotation on
250 stratified samples: 100 alternative discovery pairs,
100 NCP leaf nodes, and 50 retrieval pairs, deliberately
oversampling disagreement zones between Claude and GPT to
stress-test judge reliability.
Since sampling is non-uniform, we report population-weighted
$\kappa$~\cite{landis1977kappa} to reflect agreement on the
full evaluation set.
For alternative discovery, human--Claude weighted
$\kappa = 0.808$ (almost perfect agreement); for retrieval,
human--GPT weighted $\kappa = 0.932$ (almost perfect).
For NCP, human precision per method correlates strongly with
LLM evaluation (Pearson $r = 0.959$), and method rankings are
fully preserved: \aitd{} achieves the highest human-validated
NCP (93.3\%) across all baselines.
Results reported throughout the paper are from the first run.

All per-metric standard deviations across three independent runs
are below 0.15 on the LITE scale and 2.8\,pp on percentage-scale
metrics.

\subsection{Additional Case Studies}
\label{sec:additional-cases}

\noindent\textbf{Alternative Discovery.}
For \texttt{nextcloud/server} (self-hosted file storage), \aitd{}
returns \texttt{owncloud/core} and \texttt{pydio/cells} (P@5 = 100\%),
while the curated list (P@5 = 33\%) includes \texttt{spacedrive},
a desktop file manager serving a different purpose.

\noindent\textbf{Repository Retrieval.}
Retrieval queries are stratified into three buckets by topic-pair
intersection size: 5--15, 16--30, and 31--50 repositories.

For ``dns + vpn,'' \aitd{} scores P@5 = 0\% as its beam search
navigates to DNS-specific categories, missing the cross-domain
intersection, while Topics achieves P@5 = 100\% through direct
tag co-occurrence.
This illustrates the limitation of taxonomy-based retrieval for
queries that span orthogonal domains without a shared leaf category.

\subsection{Additional Ecosystem Finding}
\label{sec:finding4}

\observation{4}{Popularity correlates with project type.}

\noindent Repositories with $\geq$50k stars show a markedly different domain
distribution ($\chi^2 = 1{,}163$, $p < 0.001$):
28.5\% are Knowledge Resources (vs.\ 16.4\% overall)
while Library/SDK drops to 16.8\% (vs.\ 28.5\%).
The most-starred projects tend to be curated knowledge collections
(awesome-lists, tutorials, roadmaps) rather than technical libraries.
This type-level insight requires \aitd{}'s L1 artifact-type
dimension; GitHub Topics lacks a type hierarchy, and embedding
clusters group by content similarity rather than functional role.

\subsection{Additional Ablation Details}
\label{sec:ablation-details}

\noindent\textbf{Iterative refinement.}
The refinement loop's value lies in polishing dimension boundaries
and category definitions rather than in correcting gross
classification errors---hence its primary impact on structural
quality (TQ) rather than classification accuracy (NCP).

\noindent\textbf{Quality at scale.}
Beyond LITE and NCP, Sibling Coherence drops moderately on the
full 54k dataset to 70.67\% ($\Delta = -8.0$~pp from the 2k
baseline of 78.69\%).
Two factors contribute to the PG decline ($\Delta = -31.4$~pp):
first, PG assigns a binary score to each \emph{entire}
root-to-leaf path, so a single granularity misstep anywhere
zeros out the score; the 54k tree's longer paths (mean 6.1
nodes vs.\ 4.8 for 2k) amplify this penalty.
Second, maintaining consistent granularity at deeper levels is
genuinely harder with a larger, more diverse population.
Alignment-inserted nodes are collapsed before evaluation and
do not contribute.
Improving global path coherence in deeper trees is a direction
for future work.

\noindent\textbf{Level Alignment.}
Downstream task performance is unaffected by alignment: only 8 of
354 seeds show any change in alternative discovery rankings,
confirming that alignment is a non-destructive post-processing step.
On the full 54k taxonomy, alignment discovered 8 semantic ranks
(from ``Artifact Nature'' to ``Leaf-Level Differentiator''),
inserted 986 passthrough nodes, and adjusted 14,580 node depths.

\fi

\end{document}